\documentclass[12pt]{article}
\usepackage{graphicx}
\usepackage{bm}
\usepackage{dsfont}
\usepackage{amsmath}
\usepackage{amsfonts}
\usepackage{amssymb}
\usepackage{epstopdf}
\usepackage{epsfig}
\usepackage{comment}
\usepackage{array} 
\usepackage{booktabs} 
\usepackage{tabularx} 
\usepackage{setspace}
\usepackage{enumitem}
\usepackage{url}
\usepackage{rotating} 
\usepackage{booktabs}   
\usepackage{tabularx}    
\usepackage{amsmath}  
\usepackage{float}  
\usepackage{multirow}
\usepackage{color}
\usepackage{soul}
\usepackage[usenames,dvipsnames,svgnames,table]{xcolor}
\usepackage[authoryear,round]{natbib}
\usepackage{pgfplots}
\usepackage[hidelinks]{hyperref}
\hypersetup{
  colorlinks   = true, 
  urlcolor     = blue, 
  linkcolor    = blue, 
  citecolor   = blue 
}
\usepackage{appendix}

\newtheorem{corollary}{\bf Corollary}

\newtheorem{proposition}{\bf Proposition}

\newtheorem{hypothesis}{\bf Hypothesis}

\makeatletter
\newcounter{parenthypothesis}

\makeatother

\pgfplotsset{compat=1.18}

\allowdisplaybreaks

\begin{document}
\onehalfspacing

\title{Newsvendor Decisions under Stochastic and Strategic
Uncertainties: Theory and Experimental Evidence}
\author{Hang Wu \thanks{School of Economics, Zhejiang Gongshang University} \and Qin Wu\thanks{Corresponding author: The Economics Discipline Group, School of Economics, Finance and Marketing, RMIT University. Email: qin.wu@rmit.edu.au} \and Yue Liu \thanks{Shenzhen Finance Institute, School of Management and Economics, The Chinese University of Hong Kong, Shenzhen.} \and Mengmeng Shi \thanks{School of Management, Harbin Institute of Technology}}

\date{\small This version: \today}
\maketitle

\begin{abstract}
\noindent 

\noindent The rapid expansion of digital commerce platforms has amplified the strategic importance of coordinated pricing and inventory management decisions among competing retailers. Motivated by practices on leading e-commerce platforms, we analyze a sequential duopolistic newsvendor game where retailers first publicly set prices and subsequently make private inventory decisions under demand uncertainty. Our theory predicts that higher profit margins and demand uncertainty intensify price competition, while optimal inventory responses to demand uncertainty are shaped by profit margins. Laboratory evidence, however, reveals that participants are generally reluctant to compete on price, frequently coordinating on salient focal (reserve) prices, particularly in low-margin settings, and show little sensitivity to demand uncertainty in pricing. On the inventory side, participants’ order quantities are largely insensitive to chosen prices and continue to exhibit well-documented Pull-to-Center biases. These findings reveal a disconnect between pricing and inventory decisions under competition and highlight the importance of accounting for persistent behavioral tendencies in retail operations.




\end{abstract}

\section{Introduction}
\label{intro}
The growth of digital commerce platforms has elevated the importance of coordinating pricing and inventory decisions under both strategic and stochastic uncertainty. On platforms such as Amazon, sellers participating in high-profile sales events like Prime Day or Lightning Deals typically announce prices in advance while adjusting inventory closer to the sales window. Publicly posted prices act as strategic commitments that determine how demand is allocated across competing sellers. Inventory is then chosen privately after prices are observed, but before demand is realized. A similar structure is found in the fast-growing live-streaming commerce sector, on platforms such as Taobao Live and Douyin, where retailers compete for consumer attention on a daily basis. In these marketplaces, sellers often pre-announce prices ahead of scheduled broadcasts through shop listings or stream previews, then adjust inventory in response to anticipated traffic. These settings illustrate a broader operational challenge that retailers coordinate two interdependent decisions, price and inventory.


Two features of such markets are important for understanding retailers' behavior: profit margins and demand volatility. Across product categories, margins range from thin markups on everyday items to substantial markups on limited-release goods. The degree of demand uncertainty also varies considerably, from relatively stable demand for frequently purchased items to more volatile demand for new or event-based products. These two features jointly determine how retailers balance strategic and operational incentives. Higher margins increase the value of avoiding stockouts, while higher demand volatility increases inventory risk. How retailers balance these forces when price is chosen strategically and inventory is chosen under demand uncertainty is not yet well understood.
 
We study a sequential duopoly newsvendor game that captures these institutional features in a tractable way. Two retailers first set prices simultaneously. After observing each other's prices, they choose inventory levels. Each retailer faces a demand function with two parts: a deterministic component allocated based on relative prices, and a stochastic shock that captures demand uncertainty. The deterministic allocation mechanism is grounded in classic search-theoretic models of price competition (\citealp{salop1977bargains}; \citealp{varian1980model}; see \citealp{baye2006information} for a review). In these models, a fixed population of consumers allocates across sellers according to posted prices. A segment of consumers, known as ``searchers'', actively search for the lowest price and always purchase from the retailer offering the lower price. The remaining ``non-searchers'' simply buy from the first store they encounter, provided the price is below their reservation value. Hence the lower-priced seller captures the searchers whereas the higher-priced seller serves the residual non-search segment.

In line with the clearinghouse literature, we assume that the total market size is fixed. This assumption allows us to focus on how relative prices divide a given buyer pool across competing sellers and how this division interacts with inventory choices under demand uncertainty. If total demand depended on prices, price changes would affect market participation in addition to reallocating buyers across sellers. These mechanisms lead to different strategic considerations. Our analysis concentrates on the allocation mechanism because it determines the likelihood that a seller faces the high-demand or low-demand segment, and this likelihood is the object that interacts with inventory decisions in our setting. Fixing total demand allows us to study this interaction cleanly and to generate predictions that can be evaluated experimentally. Although participation responses matter in other environments, empirically, short selling horizons often feature a fixed buyer pool largely determined by external factors such as event timing or platform exposure. Hence, prices primarily influence which seller buyers choose rather than whether they enter the market. 

We characterize the subgame-perfect Nash equilibrium of the sequential pricing and inventory game. In symmetric equilibrium, sellers mix over a continuous range of prices and then choose inventory conditional on the price. Equilibrium inventory is characterized by a critical-fractile rule as in the classical newsvendor model, with the critical ratio now endogenously determined by the seller's price through its probability of capturing the high-demand segment. This structure yields clear comparative statics. Higher profit margins soften price competition and increase equilibrium inventories. Greater demand volatility shifts the mixed-strategy price distribution toward lower prices. The influence of volatility on inventory depends on the margin. At high margins, volatility raises optimal inventory, while at low margins it lowers it.

We test these predictions in a laboratory experiment that follows the two-stage game. The design varies profit margin (high vs. low) by changing unit cost and demand volatility (high vs. low) by changing the width of the demand interval. The experimental evidence is mixed compared to the predictions. On the pricing side, participants use dispersed prices, consistent with mixed-strategy competition, and prices are systematically lower in high-margin than in low-margin treatments, as predicted. However, contrary to the theoretical prediction, prices frequently cluster at the reservation value, especially in low-margin markets. This clustering suggests coordination on a salient focal point that provides payoff safety under thin margins, in line with findings from earlier studies on equilibrium selection and coordination (\citealp{dufwenberg2000price}). Demand uncertainty has little effect on pricing behavior, consistent with evidence that individuals often fail to reason backward through stochastic and strategic contingencies (\citealp{mckelvey1992experimental}; \citealp{fey1996experimental}; \citealp{binmore2002backward}). Inventory choices also deviate systematically from the equilibrium benchmark. Participants tend to overstock in low-margin conditions and understock in high-margin ones, consistent with the well-documented pull-to-center (PtC) bias (\citealp{schweitzer2000decision}; \citealp{bostian2008newsvendor}). The response to demand volatility is asymmetric. In low-margin treatments, higher uncertainty leads participants to reduce inventory, as predicted, while in high-margin treatments, inventories remain largely unresponsive. 

Our work contributes to the growing literature on joint pricing and inventory decisions under uncertainty. Prior work has established how price and inventory interact in monopolistic or simultaneous‐move settings (\citealp{petruzzi1999pricing}; \citealp{zhao2008newsvendors}; \citealp{ovchinnikov2023impact}), but little is known about how decision makers coordinate these choices when pricing is strategic and inventory is determined sequentially under demand risk. We study this problem in a tractable duopoly model in which mixed-strategy price competition determines the probability distribution over demand states, and inventory is chosen conditional on price. This structure links strategic pricing to stochastic inventory control and yields sharp comparative-static predictions that we bring to the laboratory. We document systematic departures from full coordination. Participants adjust inventory only partially to their own price and to demand volatility. These frictions indicate that the joint pricing–inventory problem imposes cognitive demands even in a simplified setting.

The rest of the paper is organized as follows. Section \ref{literature} presents a brief review of the related literature. Section \ref{theorymodel} lays out the model and characterizes the subgame perfect Nash equilibrium of the price-inventory game. Sections \ref{expdesign} and \ref{expresults} present the experimental design and results. Section \ref{conclusion} concludes.

\section{Literature Review}
\label{literature}
The classical newsvendor problem models a monopolist who selects an inventory level before demand realization (\citealp{arrow1951optimal}). In this setting, demand is exogenous and the retailer solves an individual optimization problem balancing the costs of excess inventory and lost sales, yielding the critical-fractile rule. Building on this foundation, a substantial literature examines competitive settings in which multiple retailers choose inventory levels under stochastic demand while prices remain fixed (\citealp{parlar1988game}; \citealp{wang1994three}; \citealp{lippman1997competitive}; \citealp{netessine2003centralized}). These models capture how one retailer’s stocking decision shapes the residual demand faced by others through demand substitution or reallocation, and they show that such interactions generate strategic substitutability in inventories and can lead to industry stock levels that differ markedly from a monopolist benchmark. 

A large body of research examines the joint determination of price and inventory in monopolistic newsvendor settings (\citealp{petruzzi1999pricing}; \citealp{agrawal2000impact}; \citealp{raz2006fractiles}; \citealp{aydin2008joint}; \citealp{kocabiyikouglu2011elasticity}; \citealp{kocabiyikouglu2016decision}; and \citealp{lu2013unimodality}). These models show that price and inventory must be optimized jointly. The retailer sets price to manage expected demand and margin, while inventory determines the trade-off between stockouts and excess stock. Changes in demand elasticity, cost parameters, and demand uncertainty therefore shift both the optimal price and the optimal inventory in predictable ways. Experimental studies complement this work by showing systematic behavioral deviations from normative prescriptions. Participants tend to anchor prices near costs and choose inventory levels closer to mean demand than theory predicts (\citealp{kocabiyikouglu2016decision}; \citealp{ramachandran2018multidimensional}). These findings highlight that even in the absence of competition, decision makers often struggle to integrate the two decisions in practice.

A smaller but growing literature studies environments in which retailers compete on both price and inventory. \cite{parlar2006coordinating} examine vertically related firms and show that coordinating pricing and quantity decisions can mitigate double-marginalization in markets with price-sensitive, stochastic demand. \cite{zhao2008newsvendors} develop a duopoly model in which firms choose prices and inventories simultaneously under stochastic, substitutable demand, derive conditions for the existence of pure-strategy equilibria, and characterize how equilibrium prices and quantities depend on demand and cost parameters. \cite{ovchinnikov2023impact} study an $N$-firm market in which symmetric competitors choose price and inventory simultaneously before demand realization while facing both loyal and bargain-hunting consumers. Their model shows that greater demand uncertainty generally softens price competition by reducing the incentive to undercut rivals, and that the joint decision problem often collapses into two endpoint price equilibria, reflecting the instability of intermediate prices under simultaneous choice of price and quantity. Their in-class experimental results confirm this prediction.

We extend the literature by studying a sequential price-inventory game in which prices are chosen first as strategic commitments that allocate deterministic demand, and inventories are chosen afterward under uncertainty. This structure allows the equilibrium inventory decision to take the form of a clean, price-conditional critical fractile, and yielding a tractable mixed-strategy equilibrium in prices, aligning closely with documented price dispersion observed in real retail markets (\citealp{baye2004price}). Sequential timing generates comparative statics that differ from those in simultaneous-move models. Since price is chosen before inventory, undercutting no longer requires committing to a risky stock position. Therefore, higher profit margins intensify, rather than soften, price competition, and greater demand volatility lowers prices rather than raising them. These differences highlight the importance of decision timing in competitive newsvendor environments and offer new theoretical predictions about how retailers balance price competition with inventory risk. Our framework therefore provides a clear separation between the strategic and operational components of the pricing–inventory problem and delivers testable predictions about how retailers adjust inventory after observing price competition.

Our paper is also related to a broader economics literature on price-quantity (PQ) games, where firms choose capacities (or quantities) before competing in prices. In the classical result of \cite{shubik1955comparison}, capacity choice followed by Bertrand pricing is shown to reproduce the Cournot outcome. Subsequent work introduces binding capacity constraints that generate rationing and mixed-strategy price competition (\citealp{kreps1983quantity}; \citealp{davidson1986long}). More recent contributions incorporate richer capacity or inventory constraints (\citealp{acemoglu2009price}; \citealp{montez2021all}), showing how limited stock affect price competition and market power. However, capacity in these models is typically a deterministic production limit chosen before pricing, and uncertainty in sales arises endogenously through strategic price competition rather than from exogenous demand shocks. By contrast, our newsvendor setting features inventory as a stochastic decision made after prices are observed, and retailers face classical overage and underage costs driven by demand uncertainty. These differences in both timing and the nature of uncertainty separate sequential price–inventory competition from PQ games, and they allow us to study how posted prices shape the distribution of demand states that inventory must ultimately serve.

Parallel to these theoretical advances, a rich behavioral literature has documented persistent deviations from normative predictions in inventory management. \cite{schweitzer2000decision} first documented the PtC effect, where individuals' orders gravitate toward the mean demand, with further evidence from \cite{bolton2008learning} and \cite{bostian2008newsvendor}. Subsequent studies have linked these biases to bounded rationality (\citealp{su2008bounded}; \citealp{kremer2010random}), reference dependence (\citealp{ho2010reference}), mental accounting (\citealp{chen2013effect}), prospect theory (\citealp{nagarajan2014prospect}; \citealp{long2015prospect}), and heterogeneity in risk attitudes (\citealp{bolton2012managers}; \citealp{de2013sex}). Beyond individual behavior, \cite{kirshner2019heterogeneity} examine inventory competition between newsvendors who differ in reference points and risk aversion, showing that such behavioral heterogeneity alters equilibrium order quantities and can significantly affect both firms’ profitability.

\section{Newsvendor Game and Equilibrium Solution}
\label{theorymodel}
We consider two duopolistic newsvendors, $1$ and $2$, compete in a duopolistic homogeneous product market. The game between the vendors consists of two stages. At the first stage, the vendors simultaneously and independently choose unit retailing prices from the interval $[c,r]$, where $c$ is the unit wholesale cost, and $r$ is the reservation price above which the demand is zero. The unit cost and reservation prices are exogenously determined. The demand $d_i$ for newsvendor $i=1,2$, is determined by the prices that the two newsvendors choose. If $p_i<p_j$, then newsvendor $i$ sells $d_i=d_H+\epsilon$ units of the good at price $p_i$, and newsvendor $j$ sells $d_j=d_L+\epsilon$ units of the good at price $p_j$. If $p_i=p_j$, we assume that both vendors sell $d_H+\epsilon$ and $d_L+\epsilon$ with equal probabilities. Here $d_H$ and $d_L$, with $d_H>d_L$, represent the deterministic component of demands, capturing the effect of price competition between vendors. And $\epsilon$ represents the stochastic component of demand. We assume that $\epsilon$ follows a uniform distribution $F(\cdot)$ over $[-x,x]$ with $0\le x\le d_L$. After the first stage, having observed the prices posted by both, the newsvendors choose the ordering quantities. We assume there is no switch of unmet demand between the vendors, and without loss of generality, the salvage value of leftover units is assumed to be zero. Consequently, given the chosen price $p_i$ and the realized demand $d_i$, the profit for newsvendor $i$ is $\pi_i=p_i\min{(q_i,d_i)}-cq_i$.

We solve the subgame perfect equilibrium by backward induction. Payoffs in case of price tie are omitted here and relegated to the Appendix as they bring expository complicity and have limited role in theoretical analyses for the model with continuous strategy ranges.

At stage $2$, the expected profit to newsvendor $i=1,2$ from ordering $q_i$ is: 
\begin{equation}
    E\pi_{i}\left( q_{i} \middle| \left( p_{i},p_{j} \right) \right) = p_{i}{\int_{q_{i}}^{\overset{\sim}{d} + x}q_{i}}dF\left( d_{i} \right) + p_{i}{\int_{\overset{\sim}{d} - x}^{q_{i}}d_{i}}dF\left( d_{i} \right) - cq_{i},
\end{equation}


where $\widetilde{d}=d_H$ if $p_i<p_j$ and $\widetilde{d}=d_L$ if $p_i>p_j$. Given the expected profit function, we can easily obtain the optimal ordering quantity as a function of the prices the two newsvendors chosen in stage $1$.

\begin{equation}
q_i^ * ({p_i},{p_j}) = \left\{ {\begin{array}{*{20}{c}}
{{d_H} + \left( {1 - \frac{{2c}}{{{p_i}}}} \right)x\begin{array}{*{20}{c}}
{\begin{array}{*{20}{c}}
{}
\end{array}if\begin{array}{*{20}{c}}
{{p_i} < {p_j}}
\end{array}}
\end{array}}\\
{{d_L} + \left( {1 - \frac{{2c}}{{{p_i}}}} \right)x\begin{array}{*{20}{c}}
{\begin{array}{*{20}{c}}
{}
\end{array}if\begin{array}{*{20}{c}}
{{p_i} > {p_j}}
\end{array}}
\end{array}}
\end{array}} \right.,\forall i,j \in \{ 1,2\} ,i \ne j.
\end{equation}

There is an ambiguous relationship between price and optimal inventory, which is consistent with previous observations in the theoretical literature (\citealp{raz2006fractiles}; \citealp{kocabiyikouglu2011elasticity}).
However, in our setting, lower prices imply a higher chance of winning the high demand that has a mean of $d_H$, thus driving up optimal inventory levels. On the other hand, knowing the demand distributions after the prices are posted, the optimal inventory simply follows the critical fractile solutions. Lower prices lead to lower underorder costs, hence driving the optimal inventory down.

In Stage $1$, foreseeing the optimal inventory decisions under the stochastic demand, the newsvendors choose prices to maximize expected profits. It is routine to show that for our game the discontinuous price-dependent demand structure precludes the existence of any pure-strategy equilibria for price decisions \citep{varian1980model}. In light of that, our analyses will be focusing on the symmetric equilibrium with mixed pricing strategy. Let $F_i(\cdot)$ and $F_j(\cdot)$ denote the mixed strategy in terms of cumulative probability distribution on the price range $p\in[c,r]$ for newsvendor $i$ and $j$, respectively. The expected profit for newsvendor $i$ from choosing $p_i$ is: 
\begin{equation}
    E\pi_{i}\left( {p_{i},q_{i}^{*}\left( {p_{i},p_{j}} \right)} \right) = F\left( p_{i} \right)E\pi_{i}\left( q_{i}^{*} \middle| {p_{i} > p_{j}} \right) + \left( {1 - F\left( p_{i} \right)} \right)E\pi_{i}\left( q_{i}^{*} \middle| {p_{i} < p_{j}} \right),
\end{equation}

where $E\pi_i\left(q_i^\ast\middle| p_i<p_j\right)=d_Hp_i-d_Hc-cx+\frac{c^2x}{p_i}$ is the expected profit at the optimal inventory for newsvendor $i$ when $p_i<p_j$ at the optimal ordering quantity, and $E\pi_i\left(q_i^\ast\middle| p_i>p_j\right)=d_Lp_i-d_Lc-cx+\frac{c^2x}{p_i}$ is the expected profit at the optimal inventory for newsvendor $i$ when $p_i>p_j$. Further, we define price $\widetilde{p}$ as the threshold price at which the expected profit from choosing $\widetilde{p}$ and wins the price competition is equal to choosing the reservation price $r$ and having the low demand $d\left(r\right)=d_L+\epsilon$, that is: 
\begin{equation}
d_H\widetilde{p}+\frac{c^2x}{\widetilde{p}}=d_Lr+\left(d_H-d_L\right)c+\frac{c^2x}{r}.
\end{equation}

All prices below $\tilde {p}$ will not be chosen with positive probability at the equilibrium because these prices are strictly dominated in profitability by $r$.

The subgame perfect equilibrium price strategy can be obtained from deriving the best response functions and applying the symmetry condition that both newsvendors adopt the same strategies in equilibrium. Proposition \ref{Prop:A1} characterizes the subgame perfect Nash equilibrium. For expository simplicity we relegate detailed proof of the proposition that includes equilibrium inventory decisions upon price ties to the Appendix. 

\begin{proposition} 
\label{Prop:A1}
At the symmetric subgame perfect Nash equilibrium (SPNE) for the price-inventory newsvendor game, in stage $1$ newsvendor $i=1,2$ charges a price drawn from the distribution: 

\begin{equation}
F_i^*(p) = \begin{cases} 
0, & \text{if } p \in [c, \tilde {p}) \\
1 - \frac{(r - p) \left( d_L - \frac{c^2 x}{pr} \right)}{(p - c)(d_H - d_L)}, & \text{if } p \in [\tilde{p}, r]
\end{cases};
\label{eq:pricecdf}
\end{equation}

In Stage $2$, observing the prices $(p_i,p_j)$ chosen in Stage $1$, newsvendor $i=1,2$ orders a quantity according to: 
\begin{equation}
q_i^ * ({p_i},{p_j}) = \left\{ {\begin{array}{*{20}{c}}
{{d_H} + \left( {1 - \frac{{2c}}{{{p_i}}}} \right)x\begin{array}{*{20}{c}}
{\begin{array}{*{20}{c}}
{}
\end{array}if\begin{array}{*{20}{c}}
{{p_i} < {p_j}}
\end{array}}
\end{array}}\\
{{d_L} + \left( {1 - \frac{{2c}}{{{p_i}}}} \right)x\begin{array}{*{20}{c}}
{\begin{array}{*{20}{c}}
{}
\end{array}if\begin{array}{*{20}{c}}
{{p_i} > {p_j}}
\end{array}}
\end{array}}
\end{array}} \right.,\forall i,j \in \{ 1,2\} ,i \ne j.
\label{eq:optq}
\end{equation}
\end{proposition}

Proof. See Appendix. \\
Proposition $1$ shows that at the pricing stage both vendors randomize prices according to the distribution function $F^{*}(p)$ over the range of $p\in[\tilde p,r]$, thus the equilibrium market price exhibits price dispersion. In addition, $F^{*}(p)$ is continuous over the whole support price range $[\tilde p,r]$ without any mass points, implying that the vendors persistently offer sales prices below the monopoly price $r$. 

The intuition for the mixed-strategy pricing equilibrium is straightforward.
Because the lower-priced seller captures the entire high-demand segment $d_H$ while the higher-priced seller receives only the low-demand segment $d_L$, a small price cut generates a discontinuous gain in realized demand. This discontinuity means that no price in the interior of $[\tilde p,r]$ can be a best response in pure strategies. Any such price can always be profitably undercut by an arbitrarily small amount. When both sellers randomize, however, this discontinuity is smoothed out in expected terms. For any posted price $p$, the probability of winning the high-demand segment decreases smoothly with $p$ because it depends on the opponent’s continuous price distribution. As a result, expected demand is an increasing function of own price even though realized demand is binary.

Regarding the inventory decision at stage $2$, given the outcomes of the price competition, the equilibrium inventory quantities of both newsvendors increase in their chosen prices. Moreover, for any fixed prices, the equilibrium quantity after securing the high-demand segment is greater than that after losing by a constant amount of $\left( d_{H} - d_{L} \right)$. Given the equilibrium, we obtain the following comparative static properties.


\begin{corollary}
\label{coro1}
At the equilibrium, the following hold: \\ 
 1) The lower bound of the support for the equilibrium price, $\tilde p$,decreases with $x$;\\
 2) For all $p \in [\tilde p,r] $, $F^{*}(p)$ increases with $x$; \\
 3) If $p_{i} > 2c$, $q_{i}^{*}$ increases with $x$; If $p_{i} < 2c$, $q_{i}^{*}$ decreases with $x$.
\end{corollary}

Corollary \ref{coro1} shows how demand randomness, measured by the half-width $x$ of the demand interval, shapes the subgame-perfect equilibrium. In the price stage, equilibrium prices become more competitive as $x$ increases. Two properties capture this effect. First, the lower bound of the price support,
$\tilde p$, decreases in $x$. Second, as $x$ increases, $F^{*}(p)$ increases for all prices in the support of the price distribution function. That is, for any given $p$, a larger share of the probability mass now lies below $p$.  Hence, the equilibrium distribution under a higher $x$ first-order stochastically dominates the distribution under a lower $x$.

The downward effect of demand uncertainty on equilibrium prices can be understood by examining newsvendors' trade-off in the price stage. A lower price arises the probability of obtaining the high-demand segment, whereas a higher price preserves the unit margin, $p-c$. When the half-width of the demand interval, $x$, expands, realized demand becomes harder to predict in either segment. Greater demand variation raises the expected cost of misjudging inventory as now both a stock-out and an unsold unit occur with higher probability. This heightened inventory risk makes the high-demand segment relatively more attractive than the low-demand segment, because its higher mean demand offers a larger expected sales base over which to spread any inventory error. As a result, the gain from securing the high segment increases with $x$, while the benefit of preserving a larger unit margin diminishes in relative terms. Forward-looking newsvendors therefore accept lower prices to improve their chance of capturing the high segment. In equilibrium, this behavior shifts the entire price distribution to the left and lowers the threshold price $\tilde p$.

For the second-stage inventory choice, the critical-fractile rule implies $G(q^*)=(p-c)/p$, where $G(\cdot)$ is the demand cumulative distribution. The ratio on the right hand side summarizes the trade-off between the marginal benefit of meeting an extra unit of demand and the marginal cost of holding an unsold unit. When the price posted in stage $1$ is greater than $2c$, the ratio is above $1/2$. Hence, the optimal inventory lies in the upper half of the demand distribution. As $x$ increases, a wider demand interval shifts additional probability mass to demand realizations above the previous optimal quantity, raising the expected loss from stock-outs. That is, each forgone sale unit yields a unit margin $p-c$ greater than its cost $c$, the newsvendor responds by increasing inventory as $x$ increases. In contrast, when the price is below $2c$, the ratio is below $1/2$ and the optimal quantity falls in the lower half of the demand distribution. Expanding the interval now pushes probability mass toward lower demand realizations, increasing the likelihood of overstocking. Since the unit margin $p-c$ is insufficient to cover the over-stocking cost $c$, the newsvendor reduces inventory when demand uncertainty rises. This margin-dependent response to $x$ explains why the optimal quantity $q^*$ increases in $x$ for $p>2c$ and decreases in $x$ for $p<2c$, as stated in Corollary \ref{coro1}.

\section{Experimental Design and Implementation}
\label{expdesign}

Following our theoretical model, we employ a $2\times2$ between-subject experimental design. One dimension varies the profit margin through manipulating the unit cost. Specifically, the high-margin (HM) condition has a low unit cost of $c = 3$, whereas the low-margin (LM) condition has a high unit cost of $c = 9$. The other dimension introduces two levels of demand uncertainty. Demand is drawn from a uniform distribution centred on a base level. Under the low uncertainty (LU) condition, the realized demand can deviate by at most $20$ units ($x=20$). In the high uncertainty (HU) condition, the deviation can be as large as $40$ units ($x=40$). This design yields four treatment conditions, which we label as HM\_LU, HM\_HU, LM\_LU, and LM\_HU. Each participant is randomly assigned to one of these treatments. 

We choose experimental parameters according to Proposition \ref{Prop:A1} to generate different equilibrium predictions for pricing and inventory behavior across treatments. In all treatments, the reserve price $r$ is fixed at 12. The base demand levels are set at $d_H = 100$ for the high-demand market segment and $d_L = 50$ for the low-demand segment. After the price stage, each newsvendor learns whether he or she has secured the high- or the low-demand segment but does not yet observe the realized demand. Demand realizations in each period follow a uniform distribution, represented as $d_i \sim U\left( \tilde{d} - x, \tilde{d} + x \right)$. This means, in the LU treatments ($x=20$), higher-priced participants face a demand interval of $[30,70]$ and lower-priced participants face the high-demand interval of $[80,120]$. In the HU treatments ($x=40$), these intervals widen to $[10,90]$ and $[60,140]$ for higher-priced and lower-priced participants, respectively.

Based on the parameters outlined above, we derive the equilibrium predictions for pricing and quantity decision using equations \ref{eq:pricecdf} and \ref{eq:optq} from Proposition \ref{Prop:A1}. The resulting expressions, reported in Table \ref{table:equilibrium_pred} in the Appendix, serve as the theoretical benchmark for the analysis in the next section.
 
Given that the model predicts a mixed-strategy equilibrium in pricing, precise point predictions for individual price and inventory decisions are not feasible. Therefore, we formulate our hypotheses as qualitative comparative statics, focusing on comparisons of the median and the mean of the observed choices across treatments. Our primary interest is not simply to verify whether participants' decisions precisely match the theoretical equilibrium predictions. Instead, we aim to identify persistent behavioral deviations from the equilibrium benchmarks. 

We structure our analysis around five hypotheses which are guided by theoretical benchmarks. In addition, we outline behavioral patterns that existing experimental studies suggest could pull subjects away from equilibrium. For instance, participants can have a tendency to coordinate on focal prices or to set order quantities closer to the mean demand.

We begin with pricing. Participants should draw a price from the dispersed distribution $F^*_i(p_i)$ over the interval $(\tilde p, 12)$, where $\tilde p$ is the treatment-specific threshold price and $12$ is the reserve price. If participants follow the equilibrium, no specific price point should exhibit significant clustering and prices below $\tilde p$ should be absent. This is summarized in Hypothesis 
\ref{h:pricingstrategy}. 

\begin{hypothesis} 
\label{h:pricingstrategy}
Participants select prices that are dispersed over the range of $(\tilde p,r)$, without significant clustering at specific price points. Prices below $\tilde p$ do not occur.
\end{hypothesis}

Experimental evidence on oligopolistic price competition consistently report systematic departures from theoretical predictions. In unrestricted Bertrand markets, where marginal-cost pricing is the unique Nash equilibrium, prices remain above cost and cluster at round, easily justified figures \citep{dufwenberg2000price, abbink200824}. Under capacity constraints, the Bertrand–Edgeworth model predicts a mixed-strategy price distribution, yet prices again concentrate on a narrow set of salient values rather than spanning the full equilibrium range \citep{kruse1994bertrand}. These patterns highlight the importance of focal prices, values that are both cognitively salient and strategically safe. In our experimental setup, the reserve price $r=12$ exhibits these features. Setting the reserve price guarantees at least the low-demand segment regardless of the rival’s choice, thereby eliminating strategic uncertainty. We therefore expect a noticeable concentration of price choices at $12$, contrary to the theoretical predictions of dispersed pricing over the entire interval $(\tilde p, 12)$.

Hypothesis \ref{h:pricetopmh} concerns the comparative statics of prices across treatments. The model yields a closed-form cumulative distribution $F^*_i(p_i)$ for prices and therefore a point prediction for the median price. Although the model does not pin down the average price explicitly, any leftward shift in $F^*_i(p_i)$ lowers both the median and the average. Hence, we report both statistics even though only the median is pinned down by the model. 

On the profit-margin dimension, the lower unit cost in the HM treatments reduces the indifference threshold price $\tilde p$, compared to the LM treatments. The resulting cumulative distributions of prices in the HM treatments are first-order stochastically dominated by that in the LM treatments. This implies a lower median and average price in the HM treatments. Demand uncertainty works in the same direction. The wider demand interval in the HU treatments also lowers the threshold price $\tilde p$ and shifts the price distribution leftward relative to the LU treatments. These comparative statics motivate the following hypothesis.

\begin{hypothesis}\hfill
\label{h:pricetopmh}
\begin{enumerate}[label=\textnormal{(\alph*)},
                  ref=\thehypothesis\textnormal{(\alph*)}]
    \item \label{h:pricetopmh-a}  Median and mean prices are lower in HM than in LM.  
    \item \label{h:pricetopmh-b}  Median and mean prices are lower in HU than in LU.  
\end{enumerate}
\end{hypothesis}


Moving from pricing to inventory, our next hypothesis compares inventory decisions under different market conditions. When unit cost is low (HM treatments), the unit margin $p_i-c$ comfortably exceeds the holding cost $c$. The critical-fractile rule therefore prescribes order quantities above the base levels. Specifically, in the HM treatments, the higher-priced participant is expected to order quantities above $d_L$ and the lower-priced participate should order quantities above $d_H$. In contrast, in the LM treatments where the margin is thin, higher-priced and lower-priced participants should order below $d_L$ and $d_H$, respectively. Average order quantity should thus be higher in the HM treatments than in the LM treatments.

Additionally, there is a joint effect of demand uncertainty and profit margin on inventory decisions. Corollary \ref{coro1} establishes that the optimal order quantity responds to the demand variation, $x$, in opposite directions depending on whether the posted price lies above or below $2c$. In the HM treatments, the condition $p_i \geq 2c$ always holds. Hence, the HU condition leads participants to order more than under LU condition. In contrast, the opposite inequality $p_i \leq 2c$ always applies to the LM treatments. Therefore, the HU condition leads to lower optimal quantity. We summarize these in Hypothesis \ref{h:qtod}.

\begin{hypothesis}\hfill
\label{h:qtod} 
\begin{enumerate}[label=\textnormal{(\alph*)},
                  ref=\thehypothesis\textnormal{(\alph*)}]
    \item \label{h:qtod-a}  The average order quantity is higher in HM than in LM.  
    \item \label{h:qtod-b}  Demand uncertainty works in opposite directions across margin conditions: within HM, the average order quantity is larger in HU than in LU whereas within LM it is smaller in HU than in LU.
\end{enumerate}
\end{hypothesis}

We are also interested in the interdepence between order quantity and price. The model predicts a positive relationship between the optimal order quantity $q_i^*(p_i)$ and the price chosen in the first stage, conditional on the demand segment obtained after the price stage. The logic is straightforward. Once price is set, the marginal cost of an unsold unit is the unit cost $c$, whereas the marginal benefit of having one more unit in stock equals the unit margin $p_i-c$. A higher price therefore increases the opportunity cost of a stock-out while leaving the over-stocking cost unchanged. Participants therefore find it optimal to hold more inventory when they have charged a higher price, regardless of whether they are serving the high- or the low-demand segment. Hence, the optimal quantity rises with the posted price. This leads to Hypothesis \ref{h:qwithp}.

\begin{hypothesis} 
\label{h:qwithp} 
Given the relative price outcome, the average order quantity increases with the price chosen, in every treatment.
\end{hypothesis} 

Our experiment extends the standard newsvendor framework by incorporating a competitive pricing stage in which two newsvendors simultaneously determine prices before determining inventory levels. Most experimental studies of newsvendor behavior hold price fixed and exogenous, isolating the quantity decision from any strategic pricing considerations (\citealp{schweitzer2000decision}; \citealp{bolton2008learning}; \citealp{bostian2008newsvendor}; \citealp{lurie2009timely}; \citealp{ho2010reference}; \citealp{kremer2010random}; \citealp{de2013sex}; \citealp{ren2013overconfidence}). A smaller literature allows participants to set both price and quantity but does so in a monopolistic context, abstracting from the strategic interdependence created by a competitor’s price (\citealp{ramachandran2018multidimensional}; \citealp{kocabiyikouglu2016decision}).

By pairing newsvendors in a duopoly and letting both price and quantity be endogenous, our design captures the way a competitive price choice feeds directly into the subsequent inventory decision. We assume that participants recognise this linkage. When choosing price, they anticipate that a higher unit margin will justify a larger order; when ordering, they take the first-stage price as given and adjust quantity accordingly.


Existing literature suggest that increased demand uncertainty changes inventory ordering behavior. \cite{petruzzi1999pricing} theoretically analyze a monopolistic newsvendor setting and show that as demand variance increases, the expected costs of over- or under-stocking increase, thereby altering the optimal quantity and pricing strategies. Complementing this, \cite{benzion2008decision} experimentally manipulate demand variability by keeping the mean constant but widen its spread and changing the distribution from uniform distribution to normal distribution. They find that participants deviate more from the optimal order when demand variance is high. Participants tend to rely on heuristics such as anchoring on the mean and show less accurate convergence toward optimal behavior under higher uncertainty. Their study, however, fixes the selling price and therefore cannot capture how pricing and inventory decisions interact under uncertainty. Our design allows us to test whether strategic pricing amplifies or dampens the behavioral effects documented in earlier work when competitive pricing stage is introduced.

One well-documented behavioral phenomenon in newsvendor decision-making is the Pull-to-Center (PtC) effect (\citealp{bostian2008newsvendor}; \citealp{schweitzer2000decision}; \citealp{ho2010reference}). This effect occurs when participants tend to order quantities that fall between the mean demand and the optimal order quantity. In HM treatments, participants typically underorder relative to the optimal quantity, whereas in LM treatments, they tend to overorder. These studies attribute the PtC pattern to the anchoring heuristic that decision-makers start from the unconditional mean demand and adjust only part-way toward the normative order quantity.

In our setting, the optimal quantity becomes a function of the price set $q_i^*(p_i)$. The mean demand, however, is unchanged, given the relative price outcome. We test whether the PtC pattern persists once this extra source of variation is introduced and whether it varies with price. 



\begin{hypothesis} 
\label{h:ptc} 
For every treatment and at every price level chosen in the first stage, the average order quantity lies between the unconditional mean demand and the price-contingent optimal quantity (PtC). 
\end{hypothesis}

The experiment was programmed and conducted in oTree (\citealp{chen2016otree}). In total, we ran eight sessions, with two sessions for each of the four treatments. A total of $192$ subjects, with $24$ per session, participated in the experiment. Participants were undergraduate and graduate students from Harbin Institute of Technology in Harbin, China. Each subject provided written consent and participated in only one session, and made newsvendor decisions for $50$ rounds. 

At the beginning of each session, the experimenters distributed the printed instructions and read them aloud. The instructions included numerical examples and practice questions to ensure that subjects understood how token earnings were calculated. After confirming comprehension, the experiment proceeded on computers. 

Within each session subjects were randomly assigned to fixed groups of four. These groups remained intact for the entire experiment and serve as independent observations. In every round two members of a group were randomly matched to form a duopoly, and identities were not revealed.

Each round had two stages. In stage 1 both sellers chose a price. The admissible price grid had one-decimal-place increments: $3.0$ to $12.0$ tokens in the HM treatments and $9.0$ to $12.0$ tokens in the LM treatments. After prices were posted, each seller learned whether he or she had won the high-demand segment. In stage 2 the sellers chose inventory levels. Order quantities were integers from $0$ to $120$ tokens in LM\_HU and from $0$ to $140$ tokens in LM\_LU. Unsold stock was discarded at the end of the round.

When both decisions were complete, the program drew demand, calculated profits, and displayed feedback, including their selected selling price, inventory quantity, realized demand, round profit, and the accumulated earnings. At the end of the experiment, participants completed a brief survey on demographic information, such as gender, major, school year, and prior experience with laboratory decision-making experiments.

All earnings were measured using experimental tokens. At the end of the experiment, the total accumulated token earnings from the 50 rounds were converted into Chinese Yuan (CNY). The exchange rate was set at $600$ tokens per CNY for the HM treatments and $200$ tokens per CNY for the LM treatments, ensuring that monetary incentives were comparable across the two sets of treatments. Each experimental session lasted approximately 90 minutes. On average, participants earned $44.86$ CNY, including a $20$ CNY show-up fee.

\section{Experimental Results}
\label{expresults}

\subsection{Price Decisions}
\label{pricedecision}

\begin{figure}[h]
\centering
\caption{Cumulative Probability Distributions of Observed versus NE Prices}
\includegraphics[width=\textwidth]{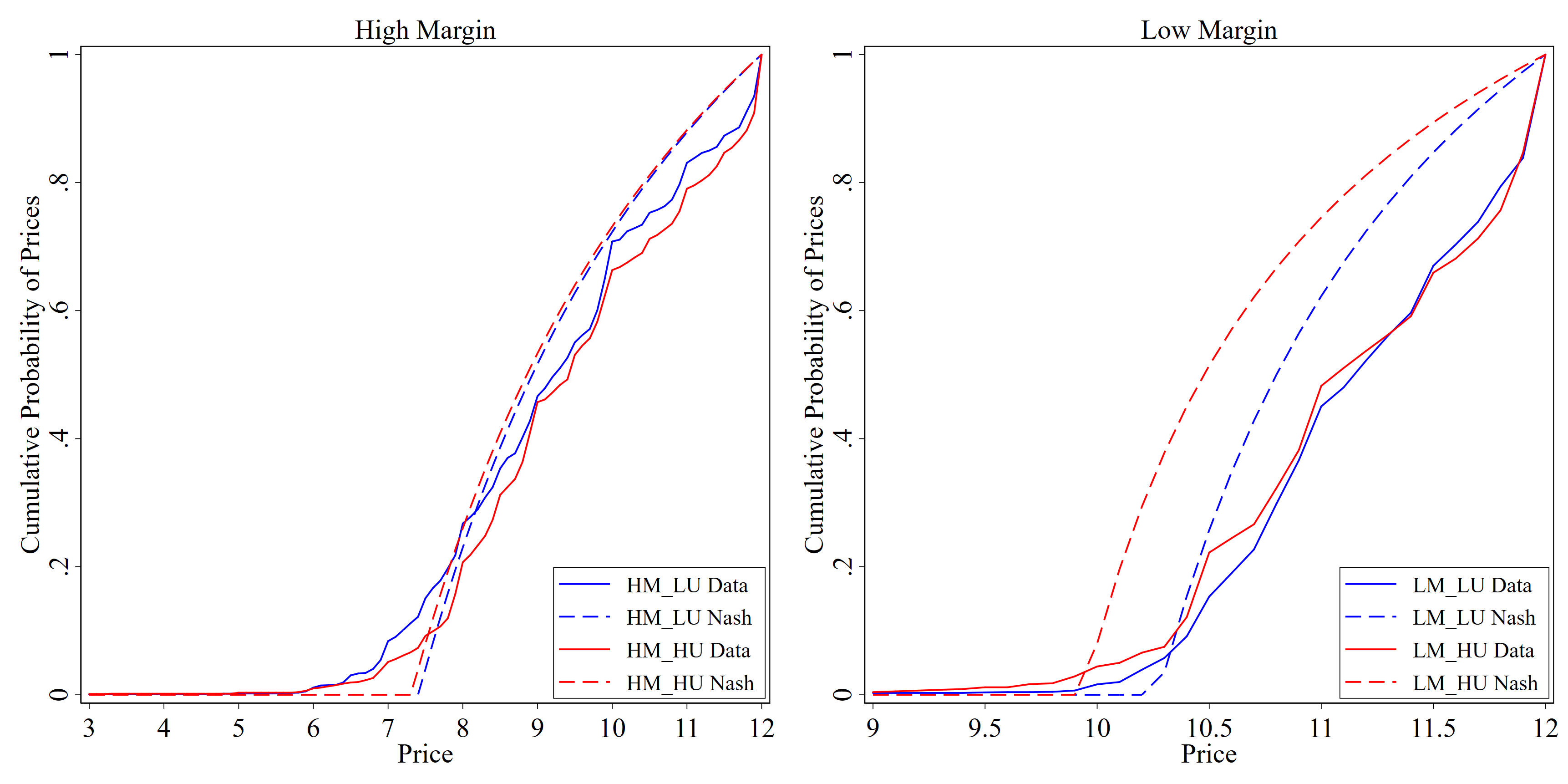}
\label{fig:pricecdf}
\end{figure}

Figure \ref{fig:pricecdf} presents the empirical cumulative distribution of prices for each treatment (solid lines) with the mixed-strategy Nash equilibrium prediction (dashed lines).The left panel displays the HM treatments and the right panel the LM treatments. 

In all treatments, prices are dispersed over the theoretical support, which is consistent with participants using mixed strategies. However, two systematic deviations from the benchmark are evident. First, there is a clear clustering at the reserve price $r=12$. In the HM\_LU and HM\_HU treatments, the reserve price accounts for $6.54\%$ and $9.13\%$ of all prices, respectively. This pattern is even more pronounced in the LM treatments, reaching $16.17\%$ in the LM\_LU treatment and $15.33\%$ in the LM\_HU treatment. The reserve price guarantees the low-demand segment and eliminates strategic uncertainty, making it a natural focal point and attracting substantial probability mass, especially when profit margins are low.
 
Another deviation appears at the lower bound of the empirical cumulative distribution. Prices below the threshold $\tilde p$ are strictly dominated by $r$. In the HM treatments, a noticeable proportion of prices fall below $\tilde p$ ($12.17\%$ in HM\_LU and $6.63\%$ in HM\_HU). This behavior is less common in the LM treatments, with $3.92\%$ and $2.88\%$ in the LM\_LU and LM\_HU treatments, respectively. This suggests that when profit margins are higher, participants are more willing to experiment with aggressive price cuts, as the cost of making an error is relatively lower. The persistent selection of dominated prices is consistent with the previous findings that participants often fail to eliminate dominated strategies, especially in environments with complex feedback and strategic uncertainty (\citealp{stahl2008level}; \citealp{camerer2004cognitive}; \citealp{eyster2005cursed}).
  
Taken together, these results partially support Hypothesis \ref{h:pricingstrategy}. Although participants randomize prices, the systematic clustering at the reserve price and the presence of dominated choices are frequent, highlighting deviations from the theoretical benchmark.



\begin{table}[tbp]
\centering
\small 
\caption{Mean of Summary Price Statistics}
\vspace{0.5em}
\begin{tabularx}{\textwidth}{l *{6}{>{\centering\arraybackslash}X}}
\toprule
Treatment &  Proportion of r & Median Price & NE Median Price & Average Price & IQR & NE IQR \\
\midrule
\multirow{2}{*}{HM\_LU} & 6.542\% & 9.188 & 8.931 & 9.330 & 2.354 & 2.097 \\
 & (0.033) & (0.721) & & (0.535) & (0.574) & \\
\multirow{2}{*}{LM\_LU} & 16.167\% & 11.167** & 10.800 & 11.214 & 0.929 & 0.765 \\
 & (0.054) & (0.198) & & (0.111) & (0.253) & \\
\multirow{2}{*}{HM\_HU} & 9.125\% & 9.313 & 8.858 & 9.531 & 2.496 & 2.141 \\
 & (0.035) & (0.592) & & (0.449) & (0.694) & \\
\multirow{2}{*}{LM\_HU} & 15.333\% & 11.200** & 10.476 & 11.1752 & 1.042 & 0.860 \\
& (0.067) & (0.305) & & (0.158) & (0.313) & \\
\bottomrule
\end{tabularx}
\parbox{\textwidth}{\footnotesize \vspace{0.5em}
\textit{Notes:} Standard deviations across fixed groups are in parentheses. Asterisks denote significant differences from Nash equilibrium using two-tailed Wilcoxon signed-rank tests, with Holm-Bonferroni correction for multiple comparisons. \\
$^*$ Significant at the 5\% level. \\
$^{**}$ Significant at the 1\% level.
}
\label{tab:observeprice}
\end{table}

Table \ref{tab:observeprice} reports summary statistics on pricing decisions for each treatment. Following the approach in \cite{cason2021experimental}'s study on price dispersion, we use the median price as the main measure of central tendency and the interquartile range (IQR) of prices as a measure of price dispersion. For each fixed group, we calculate the median price across 50 rounds. The reported treatment-level statistic is the average of these group medians, with the standard deviation across groups shown in parentheses.

Statistical comparisons using two-tailed Wilcoxon signed-rank tests indicate that the observed median prices are significantly higher than the equilibrium predictions in both LM treatments ($p<0.001$\footnote{All reported $p$ values for multiple pairwise comparisons are adjusted using the Holm-Bonferroni p-value correction to control the family-wise error rate, following the approach of \cite{cason2021experimental} (see also \cite{list2019multiple} for discussion of multiple hypothesis testing in experimental economics).} for both), but not in the HM treatments. Average prices are similar to median prices in all treatments. Furthermore, price dispersion, as measured by the IQR, is generally wider than predicted in every treatment, although the difference is insignificant. 


When comparing across profit margin conditions, we find clear treatment effects. Prices are significantly lower and more dispersed in HM treatments relative to LM treatments, in line with Hypothesis \ref{h:pricetopmh-a}. For every pairwise comparison between HM and LM treatments, the differences in median and mean prices are statistically significant ($p<0.001$, Wilcoxon rank-sum test at the group level). The difference between HM and LM treatments is also pronounced for the reserve price. The proportion of prices set at $r$ is substantially higher in LM than HM treatments   ($p<0.001$ for all comparisons).  

Turning to the effect of demand uncertainty, the results do not support Hypothesis \ref{h:pricetopmh-b}. Theoretical predictions suggest that higher uncertainty (HU) should induce lower prices, as participants would seek to offset greater stochastic risk by competing more aggressively. However, observed prices do not differ significantly between HU and LU within either profit margin condition. This absence of a significant effect is also visible in Figure \ref{fig:pricecdf}, where the empirical cumulative distributions for HU and LU nearly overlap within each panel, indicating little difference in pricing behavior in response to changes in demand uncertainty.

\begin{table}[tbp]
\centering
\small
\caption{Pooled OLS Regressions on Price and Price Change}
\vspace{0.5em}
\begin{tabularx}{\textwidth}{l *{4}{>{\centering\arraybackslash}X}}
\toprule
                    & (1)   & (2)   & (3)  & (4)  \\
                    & Price (All Rounds) & Price (2nd Half)  & Price Change (All Rounds) & Price Change (2nd Half) \\
\midrule
LM\_LU                        & $1.840^{**}$ & $1.805^{**}$ & $-0.375^{**}$ & $-0.330^{**}$ \\
                              & (0.149)      & (0.174)      & (0.046)       & (0.062)       \\[0.6ex]
HM\_HU                        & $0.135$      & $0.008$      & $-0.065$      & $0.022$       \\
                              & (0.196)      & (0.213)      & (0.064)       & (0.086)       \\[0.6ex]
LM\_HU                        & $1.794^{**}$ & $1.736^{**}$ & $-0.399^{**}$ & $-0.382^{**}$ \\
                              & (0.155)      & (0.181)      & (0.044)       & (0.061)       \\[0.6ex]
1.Lag\_Higher-P               &              &              & $-1.178^{**}$ & $-1.035^{**}$ \\
                              &              &              & (0.087)       & (0.121)       \\[0.6ex]
LM\_LU $\times$ 1.Lag\_Higher-P &              &              & $0.776^{**}$  & $0.661^{**}$  \\
                              &              &              & (0.095)       & (0.131)       \\[0.6ex]
HM\_HU $\times$ 1.Lag\_Higher-P &              &              & $0.124$       & $-0.050$      \\
                              &              &              & (0.127)       & (0.176)       \\[0.6ex]
LM\_HU $\times$ 1.Lag\_Higher-P &              &              & $0.807^{**}$  & $0.735^{**}$  \\
                              &              &              & (0.094)       & (0.125)       \\[0.6ex]
Constant  & $9.396^{**}$ & $9.472^{**}$ & $0.543^{**}$  & $0.493^{**}$  \\
                              & (0.152)      & (0.216)      & (0.049)       & (0.064)       \\[0.6ex]
\hline
Observations                  & 9,600        & 4,800        & 9,408         & 4,800         \\
Clusters                      & 48           & 48           & 48            & 48            \\
$R^2$                         & 0.3550       & 0.3694       & 0.1137        & 0.1036        \\
\bottomrule
\end{tabularx}
\parbox{\textwidth}{\footnotesize \vspace{0.5em}
\textit{Notes}: Pooled OLS regressions use robust standard errors clustered at the group level. Regressions include controls for participant gender, age range, school year, prior experience, and phase dummies (not reported). Columns (1)–(2) regress price on treatment dummies; columns (3)–(4) regress period‐to‐period price change on treatment dummies, the lagged relative price outcome dummy, and their interactions.  Columns (2) and (4) use only rounds 26–50. All post-estimation pairwise comparison p-values are Holm-adjusted. \\
$^*$ Significant at the 5\% level. \\
$^{**}$ Significant at the 1\% level.}
\label{tab:pooled_price}
\end{table}

Regression results in Table~\ref{tab:pooled_price} further support the treatment effects observed in the nonparametric analysis. Columns 1 and 2 report pooled OLS regressions of individual prices on the treatment dummies, controlling for phase (ten phases of five periods each to account for time trends) and demographic characteristics. Standard errors are clustered at the fixed group level. The results confirm that prices are systematically higher in LM than in HM treatments, consistent with Hypothesis \ref{h:pricetopmh-a}. In contrast, increasing demand uncertainty does not lead to lower prices as predicted by Hypothesis \ref{h:pricetopmh-b}. These findings are robust when the analysis is restricted to the second half of the experiment (column 2), suggesting that the absence of a demand uncertainty effect is not limited to early rounds or initial adaptation.

The evidence provides strong support for the theoretical prediction that profit margin is a primary determinant of pricing behavior. However, we do not find evidence that participants adjust prices in response to changes in demand uncertainty. This insensitivity likely arises because accounting for demand uncertainty at the pricing stage requires participants to anticipate the consequences of their price choice for uncertain future demand, a process that relies on reasoning backward from the inventory decision. The experimental literature on backward induction has documented that such anticipatory reasoning is cognitively demanding for human subjects (\citealp{mckelvey1992experimental}; \citealp{fey1996experimental}; \citealp{binmore2002backward}). 

\begin{figure}[ht]
\centering
\caption{Price Adjustment Over Time}
\includegraphics[width=\textwidth]{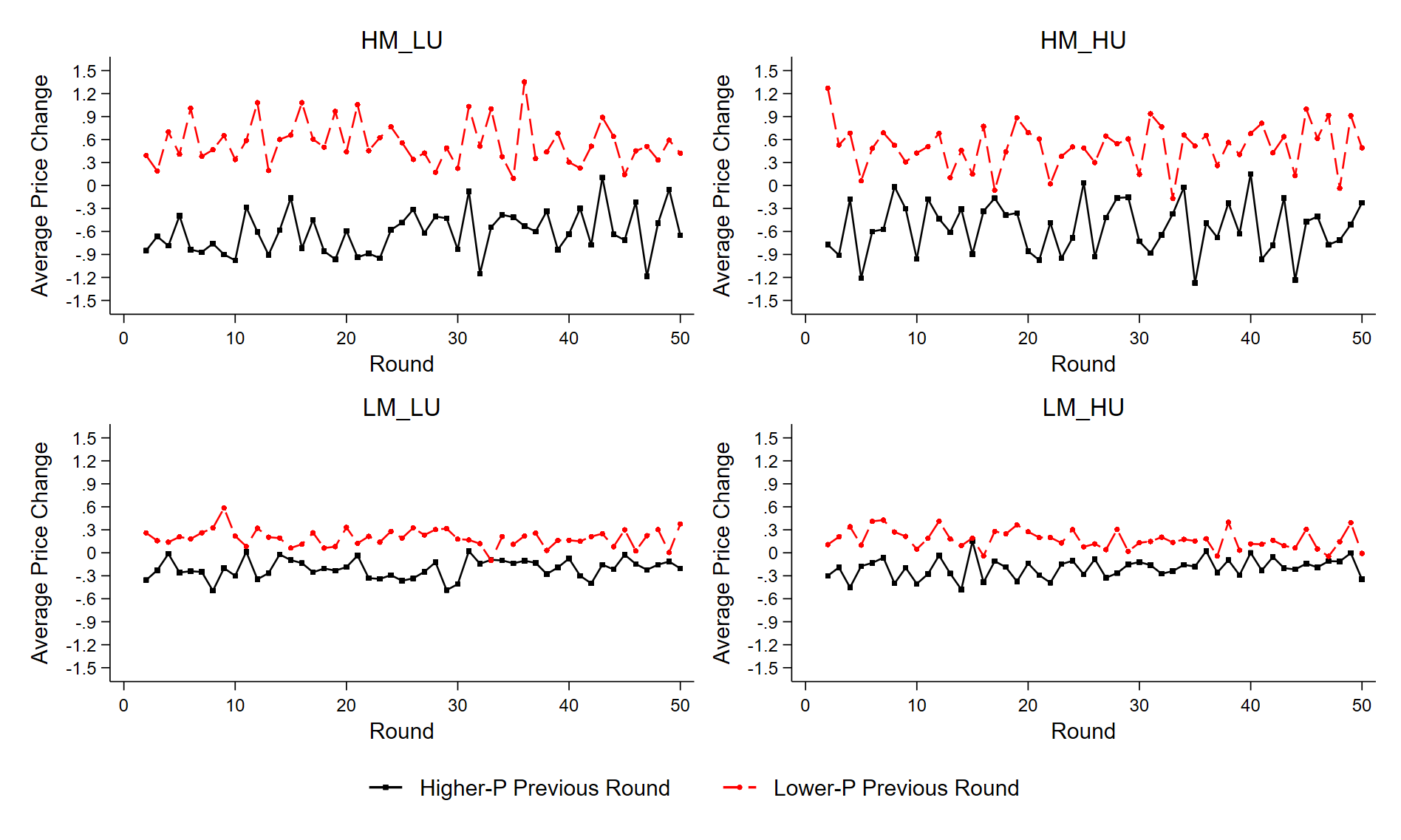}
\label{fig:pchange}
\end{figure}

We also analyze price adjustments across rounds to assess whether participants respond systematically to previous price competition outcomes. Figure \ref{fig:pchange} shows average price adjustments across rounds for each treatment, distinguishing between participants who set the lower price in the previous round and those who set the higher price. Price adjustment is measured as the difference between a participant’s current price and their price in the previous round.

The data reveal a systematic pattern of dynamic adjustment. Participants who set a lower price in the previous round tend to raise their price in the subsequent round, while those who set a higher price tend to reduce theirs. This pattern of upward adjustment by lower-priced participants and downward adjustment by higher-priced participants is present in all treatments. Moreover, the magnitude of adjustment is more pronounced in HM treatments than in LM treatments, consistent with the broader support for price choices in HM. There is little evidence that demand uncertainty (HU versus LU) affects the dynamics of price adjustment.

These patterns are supported by the pooled OLS regressions on price changes (column 3), Table~\ref{tab:pooled_price}). Explanatory variables include treatment indicators, relative price outcome from the previous round (Lag\_Higher-P), their interaction terms, phase dummies, and demographic controls, with standard errors clustered at the group level. Predicted margins highlight the magnitude of these adjustments. For participants who set the lower price in the previous round, the expected price increase in the following period ranges from $0.13$ (LM\_HU) to $0.53$ (HM\_HU) (all $p<0.001$). Conversely, those who set the higher price tend to decrease their prices by comparable magnitudes, with expected reductions from $-0.17$ (LM\_HU) to $-0.55$ (HM\_HU) (all $p<0.001$). Adjustments are systematically larger in high-margin treatments than in low-margin treatments ($p<0.001$ for all HM versus LM comparisons). These results remain robust when restricting the analysis to the second half of the experiment (column 4), with the direction and significance of effects preserved.

\subsection{Inventory Decisions}

\begin{table}[tbp]
\centering
\small 
\caption{Observed Average Quantity in Each Treatment by Price Competition Outcome}
\vspace{0.5em}
\begin{tabularx}{\textwidth}{l *{9}{>{\centering\arraybackslash}X}}
\toprule
 & \multicolumn{3}{c}{$p_i<p_j$} & \multicolumn{2}{c}{$p_i>p_j$} & \multicolumn{3}{c}{$p_i=p_j$} \\
Treatment &  $q$ & $q^*$ & $n$ & $q$ & $q^*$ & $q$ & $q^*$ & $n$ \\
\midrule
\multirow{2}{*}{HM\_LU} & 104.045** & 106.56 & 1156 & 52.350** & 56.56 & 75.463** & 93.13 & 88 \\
 & (2.064) & & & (3.236) & & (4.722) & & \\
\multirow{2}{*}{LM\_LU} & 97.897** & 86.67 & 1117 & 46.425** & 36.67 & 62.927** & 43.33 & 166 \\
 & (3.418) & & & (2.304) & & (8.293) & & \\
\multirow{2}{*}{HM\_HU} & 106.069** & 112.91 & 1148 & 51.223** & 62.91 & 75.530** & 87.91 & 104 \\
 & (4.720) & & & (5.370) & & (7.865) & & \\
\multirow{2}{*}{LM\_HU} & 90.631** & 71.27 & 1110 & 41.225** & 21.27 & 55.230** & 32.55 & 180 \\
& (6.388) & & & (5.405) & & (7.315) & & \\
\bottomrule
\end{tabularx}
\parbox{\textwidth}{\footnotesize \vspace{0.5em}
\textit{Notes:} Standard deviations across fixed groups are in parentheses. Asterisks denote significant differences from Nash equilibrium using two-tailed Wilcoxon signed-rank tests, with Holm-Bonferroni correction for multiple comparisons. \\
$^*$ Significant at the 5\% level. \\
$^{**}$ Significant at the 1\% level.
}
\label{tab:abservedaveandopt}
\end{table}

Table \ref{tab:abservedaveandopt} presents the average order quantities ($q$) and the corresponding optimal quantities ($q^*$) for each treatment, disaggregated by the relative price outcome ($p_i < p_j$, $p_i > p_j$, and $p_i = p_j$). The optimal quantities are calculated using the observed median price in each case. Standard deviations are reported across individual decisions, as inventory decisions in the second stage do not involve strategic uncertainty.

A consistent pattern emerges across all treatments. Participants tend to underorder relative to the optimal quantity when profit margins are high, and overorder when profit margins are low. This deviation from the optimal quantity is statistically significant in all cases,  with $p<0.001$ for three of the four treatments, and $p=0.0442$ for higher-priced participants in HM\_LU. This systematic bias is consistent with the well-documented PtC effect. When participants self-select prices, inventory decisions remain anchored near the mean, regardless of the optimal order.

Comparing across treatments, we find clear support for Hypothesis \ref{h:qtod-a}. Order quantities are significantly higher in HM than in LM, given the price competition outcome ($p<0.001$ for all HM versus LM comparisons). The effect of demand uncertainty is asymmetric across profit margin conditions. In HM treatments, raising uncertainty has little impact on average inventory, regardless of the price competition outcome. In contrast, under LM treatments, higher demand uncertainty leads to a significant reduction in average order quantities for both higher-priced and lower-priced participants ($p < 0.001$ for both comparisons between LM\_LU and LM\_HU). This leads to partial support to Hypothesis~\ref{h:qtod-b}. 

For the rest of the analysis of inventory decisions, we focus on cases where prices differ. This decision is supported by both theoretical and practical considerations. Theoretically, there should be no mass point over the price range, meaning that the case where $p_i=p_j$ at any price should have a probability of zero. Practically, as shown in Table \ref{tab:abservedaveandopt}, the number of observations for $p_i=p_j$ is very small across four treatments. \footnote{A coding error in the experiment caused about 2\% to 4\% of cases in four treatments to allocate both participants either to the high or low demand segments, instead of one receiving high demand and the other low demand when they set equal prices price. However, this error should not have influenced decision-making in theory, as each participant still face a 50\% chance of high demand and a 50\% chance of low demand.}

\begin{figure}[ht]
\centering
\caption{Histogram of Inventory Decisions by Price Competition Outcome and Treatment}
\includegraphics[width=\textwidth]{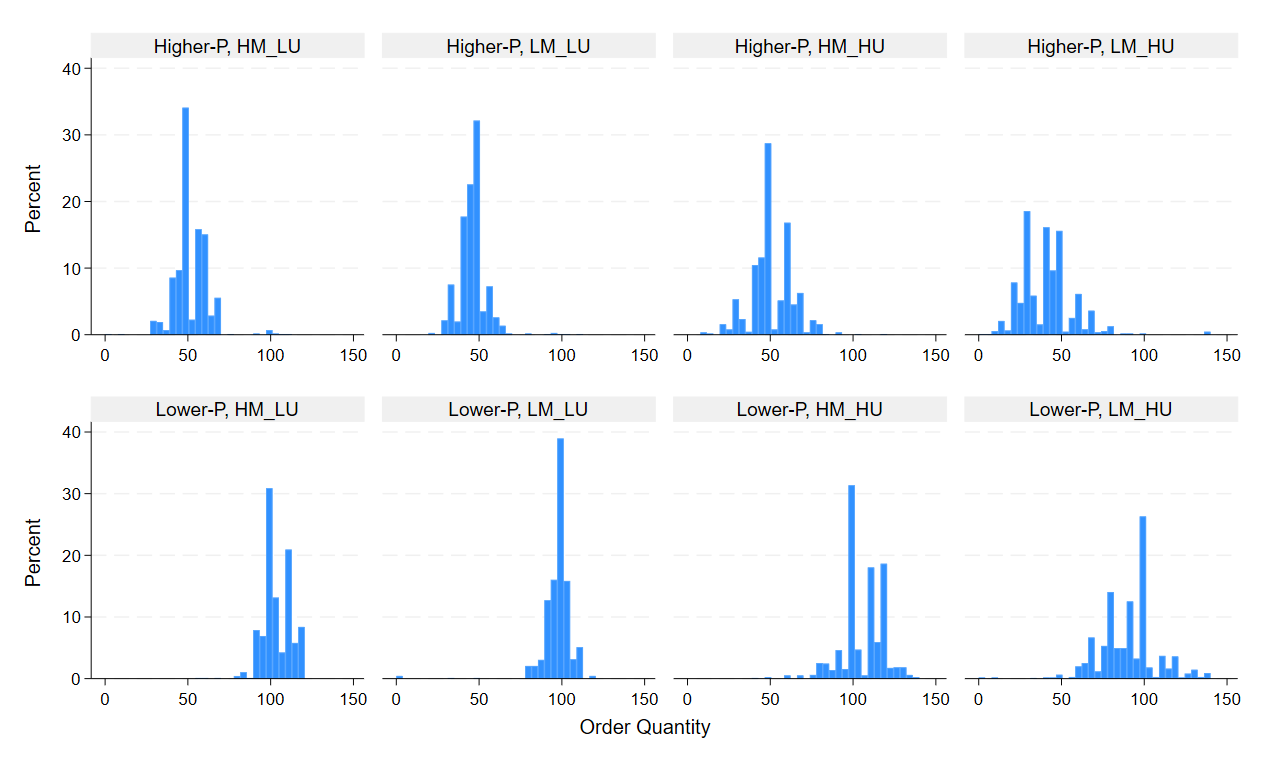}
\label{Fig_qbypriceoutcome}
\end{figure}

Figure \ref{Fig_qbypriceoutcome} displays the distribution of inventory decisions across the four treatments, with the top panels corresponding to participants who set a higher price and the bottom panels to those who set a lower price. Across all treatments and price competition outcomes, inventory choices are generally centered around the mean demand, providing further evidence of the pull-to-center effect. This tendency is apparent regardless of whether participants set higher or lower prices. However, the impact of increased demand uncertainty (HU) differs across profit margin conditions. In the HM treatments, greater uncertainty primarily increases the spread of inventory choices, with the distribution still centered near the mean. In contrast, in the LM treatments, higher demand uncertainty produces a more substantial shift. The distribution not only becomes wider but also shifts downward, especially for higher-priced participants. This pattern indicates that under low margins, participants respond to greater uncertainty by adopting more conservative inventory strategies, often ordering below the mean demand.

\begin{table}[tbp]
\centering
\small
\caption{Random Effects Regressions on Order Quantity by Price Competition Outcome}
\vspace{0.5em}
\begin{tabularx}{\textwidth}{l *{4}{>{\centering\arraybackslash}X}}
\toprule
 & (1) & (2) & (3) & (4) \\
 & Higher-Priced (All Rounds) & Lower-Priced (All Rounds) & Higher-Priced (2nd Half) & Lower-Priced (2nd Half) \\
\midrule
Price                & $0.19$      & $0.21$        & $0.01$        & $0.19$        \\
                     & (0.25)      & (0.41)        & (0.30)        & (0.33)        \\[0.6ex]
LM\_LU               & $-6.45^{**}$ & $-8.25^{**}$  & $-6.82^{**}$  & $-10.58^{**}$ \\
                     & (1.16)       & (1.82)        & (1.43)        & (2.16)        \\[0.6ex]
HM\_HU               & $-2.20$      & $1.68$        & $-1.67$       & $0.51$        \\
                     & (1.82)       & (1.33)        & (1.79)        & (1.79)        \\[0.6ex]
LM\_HU               & $-11.77^{**}$& $-14.26^{**}$ & $-13.04^{**}$ & $-17.43^{**}$ \\
                     & (1.83)       & (1.78)        & (1.96)        & (2.30)        \\[0.6ex]
Constant             & $52.93^{**}$ & $103.37^{**}$ & $50.68^{**}$  & $105.79^{**}$  \\
                     & (3.32)       & (3.77)        & (3.30)        & (3.06)        \\[0.6ex]
\hline
Observations         & 4,531        & 4,531         & 2,285         & 2,285         \\
Clusters             & 48           & 48            & 48            & 48            \\
$R^2$ (Within)  & 0.0003 & 0.0001 & 0.0001 & 0.0000 \\
$R^2$ (Between) & 0.2652 & 0.3604 & 0.2648 & 0.3421 \\
$R^2$ (Overall) & 0.1225 & 0.1953 & 0.1628 & 0.2480 \\
\bottomrule
\end{tabularx}

\parbox{\textwidth}{\footnotesize \vspace{0.5em} \textit{Notes:} Random effects regressions use robust standard errors clustered at the group level. All regressions include controls for participant gender, age range, school year and prior experience (not reported). Columns (1) and (3) restrict to higher-priced participants; columns (2) and (4) to lower-priced participants. Columns (3) and (4) use only rounds 26-50. All post-estimation pairwise comparison p-values are Holm-adjusted. \\ 
$^*$ Significant at the 5\% level. \\
$^{**}$ Significant at the 1\% level.}
\label{tab:re_orderquantity}
\end{table}

Table \ref{tab:re_orderquantity} presents the random effects regressions of order quantity by the results of price competition, separately for higher-priced participants (columns 1 and 3) and lower-priced (columns 2 and 4), both for all rounds (columns 1 and 2) and for the second half of the experiment (columns 3 and 4). The explanatory variables include the chosen price, treatment dummies, phase dummies, and demographic controls.

Consistent across all models, we find no evidence that participants systematically increase their order quantities with higher prices, as opposite to Hypothesis \ref{h:qwithp}. This suggests that, once the demand segment is realized, participants' inventory decisions are largely decoupled from their earlier pricing choices. In other words, subjects do not appear to internalize or act upon the interdependence between stage-one pricing and stage-two inventory decisions, as predicted by the theoretical model. Rather than integrating the two decisions to maximize overall profits, participants do not further optimize their inventory choice in light of the price set.


Turning to treatment effects, our results provide strong support for Hypothesis \ref{h:qtod-a}, as order quantities are significantly higher in the HM treatments relative to the LM treatments ($p<0.001$ for all HM versus LM comparisons). Regarding demand uncertainty, the regression results confirm our earlier nonparametric analysis and partially align with Hypothesis \ref{h:qtod-b}. We find no significant effect of increased demand uncertainty on order quantities in the HM treatments, suggesting that participants do not respond to the higher potential profit associated with greater demand volatility. By contrast, in the LM treatments, higher demand uncertainty leads to a significant reduction in order quantities ($p=0.009$ and $p=0.012$ for lower-priced and higher-priced participants, respectively, comparing LM\_LU and LM\_HU). These findings indicate that participants with low profit margins are more likely to treat uncertainty as a risk and respond conservatively, whereas those with high margins do not exploit uncertainty as an opportunity to expand inventory. The results are stable across the two halves of the experiment ($p=0.001$ and $p=0.014$ for lower-priced and higher-priced participants, respectively, comparing LM\_LU and LM\_HU).

Among demographic controls, we find that male participants order significantly more than female participants when setting higher prices in the second half of the experiment ($p=0.024$). No other demographic variable is statistically significant in any model. In addition, we observe that higher-priced participants exhibit a notable reduction in order quantities after the first phase, while no time pattern is evident for lower-priced participants. However, when restricting the analysis to the second half of the experiment, phase dummies do not show any significant impact in either group.

\subsection{Interdependence of Price and Inventory}

\begin{figure}[ht]
\centering
\caption{Average Order Quantity by Price: Data versus Theoretical Predictions}
\includegraphics[width=\textwidth]{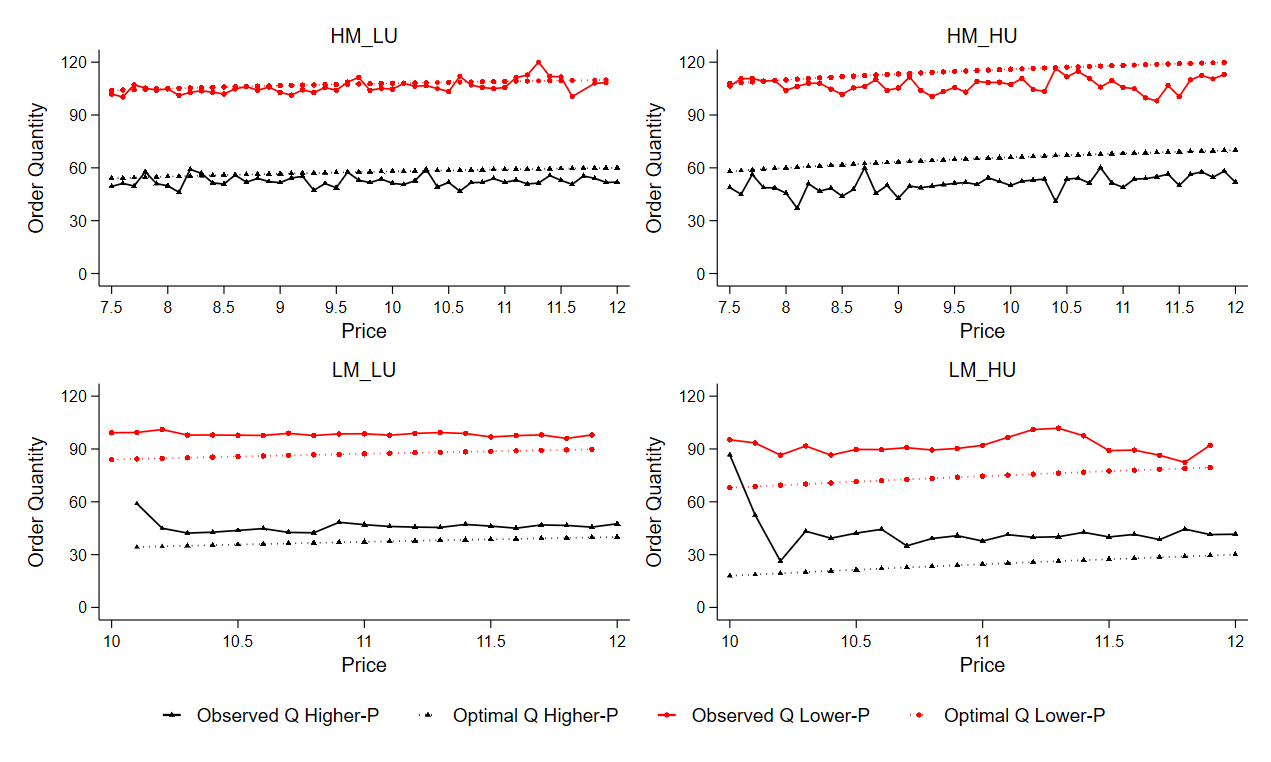}
\label{qandp}
\end{figure}

To further investigate the relationship between price and inventory choices, Figure \ref{qandp} plots average order quantities against prices, comparing observed behavior (solid lines) with theoretical benchmarks (dashed lines) in each treatment. Within each panel, red lines indicate lower-priced participants, and black lines denote higher-priced participants.

Across all treatments, participants' inventory decisions exhibit minimal responsiveness to changes in price. Average order quantities remain close to the mean demand throughout the observed price range. This reflects the persistent PtC effect, with both higher- and lower-priced participants consistently anchoring their choices near the center of the demand distribution, irrespective of margin or demand uncertainty.

This absence of systematic adjustment in inventory to price results in persistent deviations from theoretical benchmarks, with the direction of these deviations varying by treatment. Since the optimal order quantity increases with price in all treatments, participants’ failure to scale their inventories accordingly causes the gap between observed and optimal quantities to evolve differently by profit margin. In HM treatments, where average behavior is characterized by underordering, this insensitivity to price leads to increasingly pronounced underordering at higher prices. In contrast, in LM treatments, where overordering is more prevalent, the gap between actual and optimal orders narrows as price increases.

It is important to note that these patterns are observed at the average level, which may mask substantial heterogeneity in individual decision-making. Next, we examine individual-level PtC tendencies, assessing its prevalence across treatments and roles.

To systematically quantify the observed PtC bias across treatments, we adopt the standardized index developed by \cite{zhang2019meta}. However, our experimental setting fundamentally differs from previous studies considered in their meta-analysis. Prior literature typically assumes exogenously fixed prices and cost parameters, thereby yielding a fixed pair of optimal quantity and mean demand within each treatment. By contrast, in our setting, price decisions are endogenously determined, varying from round to round, which subsequently alters both the optimal inventory quantity and the corresponding mean demand faced by each participant.

In particular, the participants' strategic pricing decisions determine the demand distribution they face. Given extensive behavioral evidence that the PtC effect is primarily driven by anchoring to the mean demand, it is critical to clearly differentiate between these two distinct mean-demand anchors, $d_L$ and $d_H$, resulting from participants' competitive pricing outcomes.

Therefore, to adequately measure and interpret the PtC bias in our setting, we separately calculate the individual-level PtC index ($\alpha$) conditional on the relative price position of each participant:

\begin{equation}
\alpha_{i}^{LP}
= \frac{1}{T_i^{LP}}\sum_{t: p_{i,t}<p_{j,t}}\frac{q_{i,t}^*(p_{i,t})-q_{i,t}}{q_{i,t}-d_H},
\end{equation}

\begin{equation}
\alpha_{i}^{HP}
= \frac{1}{T_i^{HP}}\sum_{t: p_{i,t}>p_{j,t}}\frac{q_{i,t}^*(p_{i,t})-q_{i,t}}{q_{i,t}-d_L},
\end{equation}

where $q_{i,t}^*(p_i,t)$ is the round-specific optimal inventory quantity for participant $i$, $q_{i,t}$ is the actual orde quantity decision, and $d_L$ and $d_H$ are the mean demand for the higher-priced or lower-priced outcome, respectively. Thus, each conditional index precisely captures the extent to which the participant’s actual order deviates toward the conditional mean, relative to the optimal benchmark.

Next, to directly test whether the PtC effect systematically differs between price competition outcomes, we compute the within-subject PtC asymmetry statistic ($d_i$) following the methodology of \cite{zhang2019meta}:

\begin{equation}
d_i=\frac{\alpha_{i}^{LP}- \alpha_{i}^{HP}}{S_i},
\end{equation}

where $S_i$ is the within-subject pooled standard deviation of the PtC index across the two pricing outcomes. This individual-level asymmetry measure ($d_i$) explicitly quantifies differences in the strength of PtC between low-demand and high-demand rounds. A positive $d_i$ indicates a stronger PtC when higher-priced; a negative $d_i$ indicates the reverse; and $d_i$ equal to zero suggests symmetry\footnote{We calculate $S_i$, assuming independence between the higher- and lower-priced outcomes, even though actual dependence likely exists due to within-subject correlation. This assumption inflates the standard error, ensuring any identified asymmetry is robustly conservative. Later, we explicitly control for within-subject correlation through random-effects regression analysis to validate these findings.}.

In the HM treatments, the mean value of $d_i$ is negative ($-0.54$ for HM\_LU and $-0.90$ for HM\_HU), while in the LM treatments, it is positive ($0.54$ for LM\_LU and $0.33$ for LM\_HU). Wilcoxon signed-rank tests reject the null of individual symmetry ($d_i=0$) in treatments HM\_LU ($p=0.002$), HM\_HU ($p<0.001$), and LM\_LU ($p<0.001$), confirming that the PtC bias depends on the margin and relative pricing outcomes.

\begin{figure}[ht]
\centering
\caption{PtC Bias over Price Quintiles by Treatment}
\includegraphics[width=\textwidth]{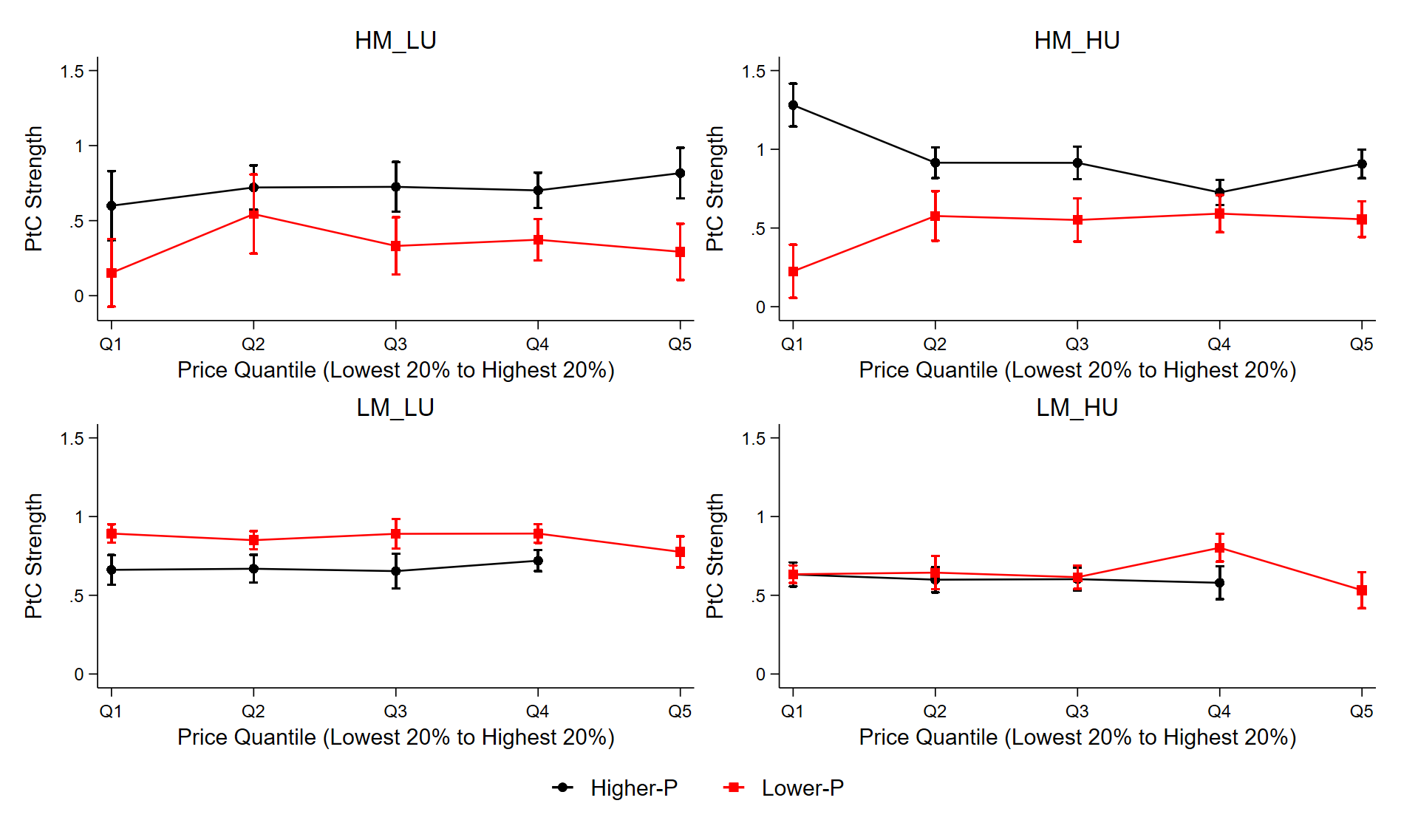}
\label{ptcandp}
\end{figure}

Figure \ref{ptcandp} further explores how PtC bias varies with chosen prices by dividing participants’ price choices into quintiles within each treatment. This quintile grouping mitigates noise from infrequently selected prices, stabilizing estimates, and allowing clearer comparisons across treatments. Each point in the figure represents the average PtC strength within a quintile, with vertical bars indicating 95 percent confidence intervals. Notably, in LM treatments, participants frequently cluster around the reservation price, limiting price variation, and hence only four distinct price quintiles are identified for higher-priced sellers.

The data reveals several persistent patterns. First, within each competitive outcome, there is no significant monotonic trend in PtC across price quintiles, indicating stable inventory bias conditional on the relative price position. Second, there is substantial asymmetry across pricing outcomes: under HM conditions, lower-priced participants consistently exhibit weaker PtC compared to their higher-priced counterparts; this relationship reverses under LM conditions, particularly in the LM\_LU treatment, where the PtC bias is significantly stronger among lower-priced participants.



\begin{table}[tbp]
\centering
\small
\caption{Random-Effects Regressions on PtC Bias}
\vspace{0.5em}
\begin{tabularx}{\textwidth}{l *{4}{>{\centering\arraybackslash}X}}
\toprule
 & \multicolumn{2}{c}{Model 1} & \multicolumn{2}{c}{Model 2} \\
 & (1) & (2) & (3) & (4) \\
 & Full Sample & 2nd Half & Full Sample & 2nd Half \\
\midrule
Price                         & 0.098        & 0.219**     &              &             \\
                              & (0.054)      & (0.042)     &              &             \\[0.3em]
Lower-P                       &              &             & -0.423*      & -0.991**    \\
                              &              &             & (0.207)      & (0.240)     \\[0.3em]
LM\_LU                        & 0.087        & -0.109      & -0.042       & -0.298*     \\
                              & (0.088)      & (0.118)     & (0.117)      & (0.121)     \\
HM\_HU                        & 0.171        & 0.240       & 0.316*       & 0.090       \\
                              & (0.150)      & (0.133)     & (0.146)      & (0.141)     \\
LM\_HU                        & -0.047       & -0.209      & -0.123       & -0.352**    \\
                              & (0.087)      & (0.116)     & (0.117)      & (0.118)     \\[0.3em]
LM\_LU $\times$ Lower-P       &              &             & 0.618**      & 1.183**     \\
                              &              &             & (0.215)      & (0.256)     \\
HM\_HU $\times$ Lower-P       &              &             & -0.259       & 0.328       \\
                              &              &             & (0.259)      & (0.300)     \\
LM\_HU $\times$ Lower-P       &              &             & 0.508*       & 1.066**     \\
                              &              &             & (0.221)      & (0.257)     \\[0.3em]
Constant                      & -0.460       & -1.711**    & 0.665**      & 0.832**     \\
                              & (0.641)      & (0.456)     & (0.120)      & (0.114)     \\[0.3em]
\hline
Observations                  & 9,043        & 4,563       & 9,043        & 4,563       \\
Clusters (groups)             & 192          & 192         & 192          & 192         \\
$R^2$ (Within)                & 0.005        & 0.029       & 0.016        & 0.047       \\
$R^2$ (Between)               & 0.012        & 0.007       & 0.007        & 0.000       \\
$R^2$ (Overall)               & 0.005        & 0.018       & 0.014        & 0.027       \\
\bottomrule
\end{tabularx}
\parbox{\textwidth}{\footnotesize
\vspace{0.5em}
\textit{Notes}: Random-effects regressions use robust standard errors clustered at the group level. All regressions include controls for participant gender, age range, school year, and prior experience (not reported). Columns (1)–(2) report PtC regressed on price and treatment indicators; columns (3)–(4) report PtC regressed on price competition outcome, treatment dummies, and their interactions. Columns (2) and (4) use only rounds 26–50. All post-estimation pairwise comparison p-values are Holm-adjusted. \\
$^*$ Significant at the 5\% level. \\
$^{**}$ Significant at the 1\% level.
}
\label{tab:ptc_reg}
\end{table}

Table \ref{tab:ptc_reg} presents regression analyses to confirm these findings. Model 1 estimates the overall effects of price and treatment indicators on PtC for both the full sample (Column 1) and the second half of the experiment (Column 2). In the latter, we find a significant positive effect of price on PtC strength ($p<0.001$). This is in line with the observation that participants do not sufficiently increase their inventory as prices rise, and therefore, amplifying PtC bias. However, there is no significant differences in PtC bias across the four treatments.

Model 2 allows us to examine how treatment effects on PtC bias differ by the pricing outcomes (higher- or lower-priced). For the full sample (Column 3), lower-priced participants in LM treatments show significantly lower PtC strength than higher-priced participants (with $p=0.078$$^\dagger$ for HM\_LU and $p<0.001$ for HM\_HU). Conversely, in the LM\_LU treatment, higher-priced participants exhibit greater PtC bias than their lower-priced counterparts ($p=0.003$) while no significant difference is observed in the LM\_HU treatment. These distinctions are robust to focusing on the second half of rounds (Column 4). The gap between higher- and lower-priced participants remains significant in both HM treatments ($p<0.001$ for both), while the reversed pattern in LM\_LU is no longer significant ($p=0.07$$^\dagger$\footnote{$^\dagger$ Unadjusted p-values for these effects are 0.039 (HM\_LU, full sample) and 0.035 (LM\_LU, second half), respectively.}). Notably, for these borderline cases, both nonparametric tests and visual inspection of the data indicate strong and consistent effects, suggesting that the loss of significance after Holm–Bonferroni correction reflects the conservative nature of the adjustment rather than an absence of underlying effect.

Our results support Hypothesis \ref{h:ptc}, confirming the presence of a robust PtC bias across all treatments. More specifically, we find that the strength and direction of this bias depend on both profit margin and the relative pricing position within each duopoly. In both HM treatments, participants secure the high demand segment exhibit significantly weaker PtC bias compared to their higher-priced counterparts, indicating a greater willingness to move inventory choices away from the mean demand when margins are high. Conversely, in the LM\_LU treatment, higher-priced participants demonstrate a stronger PtC bias than lower-priced participants, suggesting that low margins in a low-uncertainty environment exacerbate PtC bias among lower-prices participants. When demand uncertainty is high (HU), the distinction in PtC bias between higher- and lower-priced participants disappears. This suggests that increased uncertainty prompts more cautious inventory decisions, leading both groups adjust their order quantities away from the mean demand to a similar extent.

\section{Concluding Remarks}
\label{conclusion}
This paper explores the strategic interdependence between price and inventory decisions in competitive newsvendor setting, taking into account both strategic and stochastic demand uncertainties. By designing a sequential experiment that mirrors the institutional structure of modern retail platforms, where firms first compete in price and then choose inventory conditional on market segmentation, we are able to disentangle how individuals respond to both strategic and stochastic demand uncertainty.

Our results show that, while participants often adopt mixed pricing strategies as predicted by theory,participants often cluster at the reserve price, particularly in low-margin settings, reflecting a reluctance to engage in competitive price-cutting and a tendency to coordinate on salient focal points that guarantee a predictable but smaller demand segment. In contrast to theoretical predictions, we find that profit margin is the sole robust driver of pricing behavior, while demand uncertainty exerts little to no influence. This suggests that the cognitive demands of backward induction required to price for uncertainty are rarely met in practice. 

On the inventory side, profit margin again drives higher ordering, but demand uncertainty produces an asymmetric response. Specifically, while low-margin participants quickly reduce inventory as uncertainty rises, high-margin participants fail to seize the opportunity to order more. Across all treatments, inventory choices remain stubbornly anchored near the mean demand, regardless of price, leading to a persistent pull-to-center effect whose strength varies systematically with profit margin and competitive outcome. 

These findings have important managerial implications for pricing and inventory strategies in competitive retail environments. While classical models assume that firms will rationally integrate information about profit margins, demand uncertainty, and competitors’ prices, our results show that actual decision-makers often exhibit a clear disconnect between pricing and inventory choices, relying on salient focal pricing points and simple heuristics such as anchoring orders near the mean demand, rather than jointly optimizing both decisions. Managers should be aware that, even when incentives are aligned and information is transparent, cognitive complexity and behavioral coordination can prevent fully rational responses to market uncertainty. To improve operational performance, organizations may benefit from interventions that help employees explicitly connect the implications of early price choices to later inventory planning, including tools that visualize the likely demand segment and associated risks given a chosen price. Additionally, raising awareness of common behavioral tendencies, such as anchoring to reserve prices or mean demand, can be especially valuable in markets where profit margins are thin. By targeting the integration of these linked decisions and addressing context-specific biases, organizations can reduce costly mismatches between strategy and execution, and improve their adaptability in volatile marketplaces.

\newpage
\newpage
\bibliographystyle{apalike}
\bibliography{ref1}

\newpage
\appendix
\section{Appendix}

\subsection{Proof of Proposition 1.}  

We solve for the subgame-perfect Nash equilibrium by backward induction. In Stage $2$: After observing $\left( p_{i},p_{j} \right)$, the newsvendors make inventory decisions. Given the demand function:
\begin{equation}
d_{i}\left( p_{i},p_{j} \right) = 
\begin{cases}
d_{H} + \epsilon, & \text{if } p_{i} < p_{j} \\
\frac{1}{2}\left( d_{L} + \epsilon \right) + \frac{1}{2}\left( d_{H} + \epsilon \right), & \text{if } p_{i} = p_{j} \\
d_{L} + \epsilon, & \text{if } p_{i} > p_{j}
\end{cases}
\end{equation}

\noindent
\textbf{Case 1: $\boldsymbol{p_i < p_j}$.} 
Demand $d_i = d_H + \epsilon$. When $q_i \geq d_i$, newsvendor $i$ sells $d_i$ units; when $q_i < d_i$, newsvendor $i$ sells $q_i$ units. The expected profit is:
\begin{equation}
\begin{aligned}
E\pi_{i}\left( q_i \mid p_i < p_j \right) 
    &= p_i \int_{q_i}^{d_H + x} q_i  dF(d_i) + p_i \int_{d_H - x}^{q_i} d_i  dF(d_i) - cq_i \\
    &= p_i \int_{q_i}^{d_H + x} \frac{q_i}{2x}  dd_i + p_i \int_{d_H - x}^{q_i} \frac{d_i}{2x}  dd_i - cq_i \\
    &= p_i q_i \frac{d_H + x - q_i}{2x} + p_i \frac{q_i^2 - (d_H - x)^2}{4x} - cq_i
\end{aligned}
\end{equation}

The First-Order Condition (FOC) for optimal inventory is:
\begin{equation}
\frac{\partial E\pi_{i}\left( q_i \mid p_i < p_j \right)}{\partial q_i} 
    = p_i \frac{d_H + x}{2x} - \frac{p_i q_i}{x} + \frac{p_i q_i}{2x} - c = 0
\end{equation}

Optimal inventory is:
\begin{equation}
q_i^* = d_H + \left( 1 - \frac{2c}{p_i} \right) x
\end{equation}

At optimal inventory, the expected profit is:
\begin{equation}
\begin{aligned}
E\pi_{i}\left( q_i^* \mid p_i < p_j \right)
    &= p_i q_i^* \frac{d_H + x - q_i^*}{2x} + p_i \frac{q_i^{*2} - (d_H - x)^2}{4x} - cq_i^* \\
    &= d_H p_i - d_H c - c x + \frac{c^2 x}{p_i}
\end{aligned}
\end{equation}

\noindent
\textbf{Case 2: $\boldsymbol{p_i > p_j}$.} 
Demand $d_i = d_L + \epsilon$. When $q_i \geq d_i$, newsvendor $i$ sells $d_i$ units; when $q_i < d_i$, newsvendor $i$ sells $q_i$ units. The expected profit is:
\begin{equation}
\begin{aligned}
E\pi_{i}\left( q_i \mid p_i > p_j \right) 
    &= p_i \int_{q_i}^{d_L + x} q_i  dF(d_i) + p_i \int_{d_L - x}^{q_i} d_i  dF(d_i) - cq_i \\
    &= p_i \int_{q_i}^{d_L + x} \frac{q_i}{2x}  dd_i + p_i \int_{d_L - x}^{q_i} \frac{d_i}{2x}  dd_i - cq_i \\
    &= p_i q_i \frac{d_L + x - q_i}{2x} + p_i \frac{q_i^2 - (d_L - x)^2}{4x} - cq_i
\end{aligned}
\end{equation}

The First-Order Condition for profit maximization is:
\begin{equation}
\frac{\partial E\pi_{i}\left( q_i \mid p_i > p_j \right)}{\partial q_i} 
    = p_i \frac{d_L + x}{2x} - \frac{p_i q_i}{x} + \frac{p_i q_i}{2x} - c = 0
\end{equation}

Solving for optimal inventory gives:
\begin{equation}
q_i^* = d_L + \left( 1 - \frac{2c}{p_i} \right) x
\end{equation}

At optimal inventory level, the expected profit is:
\begin{equation}
\begin{aligned}
E\pi_{i}\left( q_i^* \mid p_i > p_j \right)
    &= p_i q_i^* \frac{d_L + x - q_i^*}{2x} + p_i \frac{q_i^{*2} - (d_L - x)^2}{4x} - cq_i^* \\
    &= d_L p_i - d_L c - c x + \frac{c^2 x}{p_i}
\end{aligned}
\end{equation}
\noindent

In Stage 1, foreseeing the best responses of inventory decisions to prices, the newsvendors make price decisions. The expected profit is:
\begin{equation}
\begin{aligned}
E\pi_{i}\left( p_i, q_i^*(p_i, p_j) \right) 
    &= F(p_i) E\pi_i\left( q_i^* \mid p_i > p_j \right) + \left( 1 - F(p_i) \right) E\pi_i\left( q_i^* \mid p_i < p_j \right) \\
    &= F(p_i) (d_L - d_H) (p_i - c) + d_H (p_i - c) - cx + \frac{c^2 x}{p_i}
\end{aligned}
\end{equation}

The FOC for optimal price decision is:
\begin{equation}
\frac{\partial E\pi_i\left( p_i, q_i^*(p_i, p_j) \right)}{\partial p_i} 
    = f(p_i) (d_L - d_H) (p_i - c) + F(p_i) (d_L - d_H) + d_H - \frac{c^2 x}{p_i^2} = 0
\end{equation}

Rearranging the equation gives:
\begin{equation}
F(p_i) + f(p_i) (p_i - c) = \frac{d_H - \frac{c^2 x}{p_i^2}}{d_H - d_L}
\end{equation}

As a result, we have:
\begin{equation}
\frac{d \left[ F(p_i) (p_i - c) \right]}{dp_i} = \frac{d_H - \frac{c^2 x}{p_i^2}}{d_H - d_L}
\end{equation}

Let \(k\) be a constant, the above equation implies:
\begin{equation}
F(p_i) (p_i - c) = \frac{d_H}{d_H - d_L} p_i + \frac{c^2 x}{p_i (d_H - d_L)} - k
\end{equation}

Next, we apply \(F(\tilde{p}) = 0\) to solve for \(k\). Price \(\tilde{p}\) is the threshold at which the expected payoff equals choosing monopoly price \(r\) and facing demand \(d_i(r) = d_L + \epsilon\):
\begin{equation}
d_H \tilde{p} - d_H c - cx + \frac{c^2 x}{\tilde{p}} = d_L r - d_L c - cx + \frac{c^2 x}{r}
\end{equation}

Simplifying gives:
\begin{equation}
d_H \tilde{p} + \frac{c^2 x}{\tilde{p}} = d_L r + (d_H - d_L) c + \frac{c^2 x}{r}
\end{equation}

Rearranging the equation, we obtain:
\begin{equation}
\frac{d_H}{d_H - d_L} \tilde{p}_i + \frac{c^2 x}{\tilde{p}_i (d_H - d_L)} 
    = \frac{d_L r}{d_H - d_L} + c + \frac{c^2 x}{r (d_H - d_L)}
\end{equation}

From \(F(\tilde{p}_i) = 0\):
\begin{equation}
\frac{d_H}{d_H - d_L} \tilde{p}_i + \frac{c^2 x}{\tilde{p}_i (d_H - d_L)} - k = 0
\end{equation}

As a result:
\begin{equation}
k = \frac{d_L r}{d_H - d_L} + c + \frac{c^2 x}{r (d_H - d_L)}
\end{equation}

Given \(k\), we have:
\begin{equation}
F(p_i) (p_i - c) = \frac{d_H}{d_H - d_L} p_i + \frac{c^2 x}{p_i (d_H - d_L)} 
    - \left( \frac{d_L r}{d_H - d_L} + c + \frac{c^2 x}{r (d_H - d_L)} \right)
\end{equation}

Finally, we obtain the equilibrium price distribution function:
\begin{equation}
\begin{aligned}
F^*(p_i) 
    &= \frac{1}{(p_i - c)(d_H - d_L)} 
        \left[ d_H p_i + \frac{c^2 x}{p_i} - d_L r - (d_H - d_L)c - \frac{c^2 x}{r} \right] \\
    &= \frac{1}{(p_i - c)(d_H - d_L)} 
        \left[ (d_H - d_L)(p_i - c) - d_L (p_i - r) + \left( \frac{1}{p_i} - \frac{1}{r} \right) c^2 x \right] \\
    &= 1 - \frac{1}{(p_i - c)(d_H - d_L)} 
        \left[ d_L (p_i - r) - \left( \frac{r - p_i}{p_i r} \right) c^2 x \right] \\
    &= 1 - \frac{(r - p_i) \left( d_L - \frac{c^2 x}{p_i r} \right)}{(p_i - c)(d_H - d_L)}
\end{aligned}
\end{equation}

This completes the proof.

\subsection{Optimal inventory solutions at price ties}
When \(p_i = p_j\), we discuss situations with and without overlapping between the high- and low-demand distributions. In each situation, we solve for the optimal inventory \(q_i\).

\noindent
\textbf{Situation 1: Without overlapping.}
If \(p_i = p_j\), the expected profit of newsvendor \(i\) is given by:
\begin{equation}
E\pi_i\left( q_i \mid p_i = p_j \right) = 
\begin{cases}
\frac{1}{2}p_i q_i + \frac{1}{2}p_i \min\left( q_i, d_L + \epsilon \right) - c q_i, & \text{if } q_i \in [d_L - x, d_L + x] \\[1ex]
\frac{1}{2}p_i q_i + \frac{1}{2}p_i d_L - c q_i, & \text{if } q_i \in [d_L + x, d_H - x] \\[1ex]
\frac{1}{2}p_i \min\left( q_i, d_H + \epsilon \right) + \frac{1}{2}p_i d_L - c q_i, & \text{if } q_i \in (d_H - x, d_H + x]
\end{cases}
\end{equation}

(1) If $q_{i} \in \left\lbrack d_{L} - x,d_{L} + x \right)$, $
E\pi_{i} = - \frac{p_{i}}{8x}q_{i}^{2} + \left( {\frac{p_{i}\left( d_{L} + 3x \right)}{4x} - c} \right)q_{i} - \frac{{p_{i}\left( {d_{L} - x} \right)}^{2}}{8x}$. Given the characteristics of the quadratic function, when $p_{i} < 2c$, $E\pi_{i}$ initially increases in $q_i$ over $\left\lbrack d_{L} - x,d_{L} + \left( {3 - \frac{4c}{p_{i}}} \right)x \right\rbrack$, reaching its maximum at $q_{i} = d_{L} + \left( {3 - \frac{4c}{p_{i}}} \right)x$, and subsequently decreases as $q_i$ increases. When $p_{i} > 2c$, $E\pi_{i}$ exhibits an increasing trend in $q_i$ over $\left\lbrack d_{L} - x,d_{L} + x \right)$.

(2) If $q_{i} \in \left\lbrack d_{L} + x,d_{H} - x \right)$, $E\pi_{i} = \frac{1}{2}p_{i}q_{i} + \frac{1}{2}p_{i}d_{L} - cq_{i}$. Following the characteristics of the linear function, when $p_{i} < 2c$, $E\pi_{i}$ rises with increasing $q_i$ over $\left\lbrack d_{L} + x,d_{H} - x \right\rbrack$, while it declines with increasing $q_i$ when $p_{i} > 2c$.

(3) If $q_{i} \in \left( d_{H} - x,d_{H} + x \right\rbrack$, $
E\pi_{i} = - \frac{p_{i}}{8x}q_{i}^{2} + \left( {\frac{p_{i}\left( {d_{H} + x} \right)}{4x} - c} \right)q_{i} + \frac{1}{2}p_{i}d_{L} - \frac{{p_{i}\left( {d_{H} - x} \right)}^{2}}{8x}$. According to the properties of the quadratic function, when $p_{i} < 2c$, $E\pi_{i}$ decreases in $q_i$ over $\left( d_{H} - x,d_{H} + x \right\rbrack$. When $p_{i} > 2c$, $E\pi_{i}$ initially increases in $q_i$ over $
\left\lbrack d_{H} - x,d_{H} + \left( 1 - \frac{4c}{p_{i}} \right)x \right\rbrack$, reaching its maximum at $q_{i} = d_{H} + \left( {1 - \frac{4c}{p_{i}}} \right)x$, and then decreases in $q_i$.

Combining the outcomes from (1) to (3), if $p_i=p_j$, the optimal inventory decision for newsvendor $i$ is:
\begin{equation}
    q_{i}^{*} = 
    \begin{cases}
        d_{L} + \left( 3 - \frac{4c}{p_{i}} \right)x, & \text{if } p_{i} < 2c \\
        d_{H} + \left( 1 - \frac{4c}{p_{i}} \right)x, & \text{if } p_{i} > 2c
    \end{cases}
    \quad\text{ and } \quad q_{i}^{*} \in \left[ d_{L} + x, d_{H} - x \right] \text{ if } p_{i} = 2c.
\end{equation}

\noindent
\textbf{Situation 2: with overlapping.} If $p_i=p_j$, the expected profit of newsvendor $i$ is given by:
\begin{equation}
E\pi_{i}\left( q_{i} \middle| p_{i} = p_{j} \right) = -cq_{i} + \begin{cases}
\dfrac{1}{2}p_{i}q_{i} + \dfrac{1}{2}p_{i} \cdot \min\left\{ q_{i},d_{L} + \epsilon \right\}, & 
    \text{if } q_{i} \in [d_{L} - x, d_{H} - x] \\[10pt]
\dfrac{1}{2}p_{i} \cdot \min\left\{ q_{i},d_{H} + \epsilon \right\} \\ \quad + \dfrac{1}{2}p_{i} \cdot \min\left\{ q_{i},d_{L} + \epsilon \right\}, & 
    \text{if } q_{i} \in [d_{H} - x, d_{L} + x] \\[10pt]
\dfrac{1}{2}p_{i} \cdot \min\left\{ q_{i},d_{H} + \epsilon \right\} + \dfrac{1}{2}p_{i}d_{L}, & 
    \text{if } q_{i} \in [d_{L} + x, d_{H} + x]
\end{cases}
\end{equation}

(1) If $q_{i} \in \left\lbrack d_{L} - x,d_{H} - x \right)$, $E\pi_{i} = - \frac{p_{i}}{8x}q_{i}^{2} + \left( {\frac{p_{i}\left( d_{L} + 3x \right)}{4x} - c} \right)q_{i} - \frac{{p_{i}\left( {d_{L} - x} \right)}^{2}}{8x}$. Given the characteristics of the quadratic function, when $p_{i} < \frac{4cx}{4x - \left( d_{H} - d_{L} \right)}$, $E\pi_{i}$ initially increases in $q_i$ over $\left\lbrack d_{L} - x,d_{L} + \left( {3 - \frac{4c}{p_{i}}} \right)x \right\rbrack$, reaching its maximum at $
q_{i} = d_{L} + \left( {3 - \frac{4c}{p_{i}}} \right)x$, and subsequently decreases as $q_i$ increases. When $p_{i} > \frac{4cx}{4x - \left( d_{H} - d_{L} \right)}$, $E\pi_{i}$ exhibits an increasing trend in $q_i$ over $\left\lbrack d_{L} - x,d_{H} - x \right)$.

(2) If $q_{i} \in \left\lbrack d_{H} - x,d_{L} + x \right)$, $E\pi_{i} = - \frac{p_{i}}{4x}q_{i}^{2} + \left( {\frac{p_{i}\left( {d_{H} + d_{L} + 2x} \right)}{4x} - c} \right)q_{i} - \frac{p_{i}\left\lbrack \left( {d_{H} - x} \right)^{2} + \left( {d_{L} - x} \right)^{2} \right\rbrack}{8x}$. Following the characteristics of the quadratic function, when $p_{i} < \frac{4cx}{4x - \left( d_{H} - d_{L} \right)}$, $E\pi_{i}$ decreases with increasing $q_i$ over $\left\lbrack d_{H} - x,d_{L} + x \right\rbrack$. When $\frac{4cx}{4x - \left( d_{H} - d_{L} \right)} < p_{i} < \frac{4cx}{d_{H} - d_{L}}$, $E\pi_{i}$ initially increases in $q_i$ over $\left\lbrack d_{H} - x,q_{i} = \frac{1}{2}\left( {d_{H} + d_{L}} \right) + \left( {1 - \frac{2c}{p_{i}}} \right)x \right\rbrack$, reaching its maximum at $q_{i} = \frac{1}{2}\left( {d_{H} + d_{L}} \right) + \left( {1 - \frac{2c}{p_{i}}} \right)x$, and then decreases in $q_i$. When $
p_{i} > \frac{4cx}{d_{H} - d_{L}}$, $E\pi_{i}$ increases with increasing $q_i$ over $
\left\lbrack d_{H} - x,d_{L} + x \right\rbrack$.

(3) If $q_{i} \in \left( d_{L} + x,d_{H} + x \right\rbrack$, $
E\pi_{i} = - \frac{p_{i}}{8x}q_{i}^{2} + \left( {\frac{p_{i}\left( {d_{H} + x} \right)}{4x} - c} \right)q_{i} + \frac{1}{2}p_{i}d_{L} - \frac{{p_{i}\left( {d_{H} - x} \right)}^{2}}{8x}$. According to the properties of the quadratic function, when $
p_{i} < \frac{4cx}{d_{H} - d_{L}}$, $E\pi_{i}$ decreases in $q_i$ over $\left( d_{L} + x,d_{H} + x \right\rbrack$. When $p_{i} > \frac{4cx}{d_{H} - d_{L}}$, $E\pi_{i}$ initially increases in $q_i$ over $\left\lbrack d_{L} + x,d_{H} + \left( 1 - \frac{4c}{p_{i}} \right)x \right\rbrack$, reaching its maximum at $q_{i} = d_{H} + \left( {1 - \frac{4c}{p_{i}}} \right)x$, and then decreases in $q_i$.

Combining the outcomes from (1) to (3), if $p_i=p_j$, the optimal inventory decision for newsvendor $i$ is:
\begin{equation}
q_i^* = \begin{cases}
d_L + \left( 3 - \frac{4c}{p_i} \right)x, & 
\text{if } p_i < \frac{4cx}{4x - (d_H - d_L)} \\
\frac{1}{2} (d_H + d_L) + \left( 1 - \frac{2c}{p_i} \right)x, & 
\text{if } \frac{4cx}{4x - (d_H - d_L)} \le p_i \le \frac{4cx}{d_H - d_L} \\
d_H + \left( 1 - \frac{4c}{p_i} \right)x, & 
\text{if } p_i > \frac{4cx}{4x - (d_H - d_L)}
\end{cases}
\end{equation}

\vspace{0.5em}
\subsection{Equilibrium Predictions by Experimental Treatment}
\vspace{-1.5em}
\begin{table}[H]
\centering
\footnotesize
\caption{Equilibrium Predictions by Experimental Treatment}
\vspace{0.3em}
\begin{tabularx}{\linewidth}{@{}>{\raggedright\arraybackslash}p{2.3cm}>{\raggedright\arraybackslash}X>{\raggedright\arraybackslash}X@{}} 
\toprule
\textbf{Treatment}  & \textbf{\qquad \qquad \qquad \qquad $F_i^*(p_i)$} & \textbf{\qquad \qquad \qquad $q_i^*(p_i)$} \\
\midrule
HM\_LU 
& 
$\begin{cases}
0,&\text{if } p_i \in [3.0,\,7.4] \\[4pt]
1 - \dfrac{(12-p_i)(10p_i-3)}{10p_i(p_i-3)}, &\text{if } p_i \in [7.5,\,12.0]
\end{cases}$
& 
$\begin{cases}
\qquad 120 - \dfrac{120}{p_i}, & \text{if } p_i < p_j \\[9pt]
\qquad 120 - \dfrac{240}{p_i}, & \text{if } p_i = p_j \\[9pt]
\qquad 70 - \dfrac{120}{p_i}, & \text{if } p_i > p_j
\end{cases}$
\\
\addlinespace[0.5ex]
\hline
\addlinespace[0.5ex]
LM\_LU 
& 
$\begin{cases}
0, & \text{if } p_i \in [9.0,\,10.2] \\[4pt]
1 - \dfrac{(12-p_i)(10p_i-27)}{10p_i(p_i-9)}, & \text{if } p_i \in [10.3,\,12.0]
\end{cases}$
& 
$\begin{cases}
\qquad 120 - \dfrac{360}{p_i}, & \text{if } p_i < p_j \\[9pt]
\qquad 110 - \dfrac{720}{p_i}, & \text{if } p_i = p_j \\[9pt]
\qquad 70 - \dfrac{360}{p_i}, & \text{if } p_i > p_j
\end{cases}$
\\
\addlinespace[0.5ex]
\hline
\addlinespace[0.5ex]
HM\_HU 
& 
$\begin{cases}
0, & \text{if } p_i \in [3.0,\,7.3] \\[4pt]
1 - \dfrac{(12-p_i)(10p_i-6)}{10p_i(p_i-3)}, & \text{if } p_i \in [7.4,\,12.0]
\end{cases}$
& 
$\begin{cases}
\qquad 140 - \dfrac{240}{p_i}, & \text{if } p_i < p_j \\[9pt]
\qquad 115 - \dfrac{240}{p_i}, & \text{if } p_i = p_j \in [7.4,\,9.6] \\[9pt]
\qquad 140 - \dfrac{480}{p_i}, & \text{if } p_i = p_j \in [9.7,\,12.0] \\[9pt]
\qquad 90 - \dfrac{240}{p_i}, & \text{if } p_i > p_j
\end{cases}$
\\
\addlinespace[0.5ex]
\hline
\addlinespace[0.5ex]
LM\_HU 
& 
$\begin{cases}
0, & \text{if } p_i \in [9.0,\,9.9] \\[4pt]
1 - \dfrac{(12-p_i)(10p_i-54)}{10p_i(p_i-9)}, & \text{if } p_i \in [10.0,\,12.0]
\end{cases}$
& 
$\begin{cases}
\qquad 140 - \dfrac{720}{p_i}, & \text{if } p_i < p_j \\[9pt]
\qquad 170 - \dfrac{1440}{p_i}, & \text{if } p_i = p_j \\[9pt]
\qquad 90 - \dfrac{720}{p_i}, & \text{if } p_i > p_j
\end{cases}$
\\
\bottomrule
\end{tabularx}
\label{table:equilibrium_pred}
\end{table}

\subsection{Translated Sample Instructions (HM-LU treatment)}
Thank you for participating in our economic management decision-making experiment. Please read these instructions carefully to ensure you understand the decision-making process.

In this session, you will receive a $20$ RMB participation fee, which is not affected by your decisions in the experiment. Additionally, you will have the opportunity to earn experimental currency units in each round to generate extra income. At the end of the experiment, your accumulated experimental currency units will be converted to RMB at an exchange rate of $600:1$ (every $600$ experimental currency units equal $1$ RMB). Your final earnings will be calculated as: 
\begin{equation*}
\begin{aligned}
\text{20 RMB Show-up fee}  
& + \left( \text{Total experimental currency units over 50 rounds} \div 600 \right)
\end{aligned}
\end{equation*}
Your decisions may result in gains or losses in experimental currency units. However, regardless of your final experimental currency balance, you will always receive at least the $20$ RMB participation fee. Your total earnings will be paid to you via WeChat transfer at the conclusion of the experiment.

Please note that this is a strictly controlled scientific experiment. Chatting, making loud noises, looking at others’ screens, or any other disruptive behaviors are prohibited. If you have questions during the experiment, please raise your hand, and our staff will assist you promptly. Thank you for your cooperation!

\textbf{Experimental Decision-Making Process:}

\textbf{(1)} This experiment consists of $50$ identical and independent rounds of market decision-making. Before the experiment begins, the system will randomly assign you and three other participants in this session to form a fixed large group, which will remain unchanged throughout the entire experiment. At the start of each subsequent round, the system will randomly pair you with one other member from your large group to form a small group. At the start of each round, the system will randomly pair you with another member from your large group to form a small group. You and your paired member will compete in the same market by selling identical products.

For example, if you are grouped with Participants $8$, $14$, and $21$ in your large group, the system will randomly pair you with Participant $8$, $14$, or $21$ at the start of each new round.

\textbf{(2)} In each round, the procurement cost of each product is $3.0$ experimental currency units. You must first decide the selling price of the product. You may choose any value within the range $\{3.0,3.1,3.2,\dots,12.0\}$ as the selling price.

Note: Your selected price must not be lower than $3.0$ or higher than $12.0$, and you should submit your price within $20$ seconds.

\textbf{(3)} After both you and your opponent submit your selling prices, the computer screen will display your price and your opponent’s price. The market demand is randomized and will be allocated between you and your opponent based on the following rules:
\begin{itemize}
    \item If your selling price is lower than your opponent’s, you will receive high demand. The system will randomly select an integer from the range $\{80,81,\dots,120\}$ (all values equally likely) as your actual demand.
    \item If your selling price is higher than your opponent’s, you will receive low demand. The system will randomly select an integer from the range $\{30,31,\dots,70\}$(all values equally likely) as your actual demand.
    \item If your selling price equals your opponent’s, there is a 50\% chance you will receive high demand (range: $\{80,81,\dots,120\}$) while your opponent receives low demand (range: $\{30,31,\dots,70\}$), and a 50\% chance the reverse will occur (you receive low demand, your opponent receives high demand).
    
    Based on your opponent’s selling price and the market demand allocation rules, you must now decide how many units to procure from the wholesaler. You may select any integer within the range $\{0,1,2,\dots,150\}$ as your procurement quantity. At the end of each round, unsold goods will be automatically cleared (discarded without compensation).
    
    Note: Your procurement quantity must not be less than $0$ or exceed $150$, and you should submit your decision within $20$ seconds.
\end{itemize}

\textbf{(4)} After both you and your opponent submit your procurement quantities, the system will allocate market demand according to the rules above and automatically calculate your earnings for the round. The calculation rules are as follows:

\begin{itemize}
    \item If your procurement quantity is less than or equal to your actual demand, all procured goods will be sold. Your experimental currency earnings for the round are:
    \begin{equation*}
    \begin{aligned}
    \text{Earnings} &= \left( \text{Procurement Quantity} \times \text{Your Selling Price} \right) \\
    & - \left( \text{Procurement Quantity} \times 3.0 \right)
    \end{aligned}
    \end{equation*}   

     \item If your procurement quantity exceeds your actual demand, only a portion of goods will be sold. Your experimental currency earnings for the round are:
    \begin{equation*}
    \begin{aligned}
    \text{Earnings} &= \left( \text{Actual Demand} \times \text{Your Selling Price} \right) \\
    & - \left( \text{Procurement Quantity} \times 3.0 \right)
    \end{aligned}
    \end{equation*}
\end{itemize}

\textbf{(5)} At the end of each round, the screen will display: 

Your and your opponent’s decisions and earnings for the current round. 

Historical data from all previous rounds, including: your selling price, your procurement quantity, your actual demand, your current round earnings, your cumulative total earnings.

\textbf{The following is an example:}

The procurement cost is $3.0$ experimental currency units per unit. Suppose:

Your selling price in this round is $11.5$ experimental currency units per unit;

Your opponent’s selling price is $10.2$ experimental currency units per unit.

Since your price exceeds your opponent’s price, you will receive low demand. The system will randomly select an integer from the range $\{30,31,\dots,70\}$ (all values equally likely) as your actual demand.

Assume you choose a procurement quantity of $62$ units. After both you and your opponent submit your procurement decisions, the system randomly assigns your actual demand as $59$ units.

Your earnings for this round would be: $59\times11.5-62\times3.0=492.5$ experimental currency units.

\end{document}